\pdfoutput=1
\documentclass[11pt,twoside,a4paper,cmspaper,final,collab]{cms-tdr}
\def\svnVersion{1318c0e}\def\svnDate{2026/07/21}\def\cmsCernNoTag{CERN-EP-2025-296}\def\cmsCernDate{\today}\def\cmsMessage{Published in the Journal of High Energy Physics as \href{http://dx.doi.org/10.1007/JHEP07(2026)102}{\doi{10.1007/JHEP07(2026)102}.}}
\begin{document}\cmsNoteHeader{EXO-24-020}

\newcommand{\centerstack}[2][c]{%
  \raisebox{-0.25\height}{\shortstack[#1]{#2}}%
}

\newcommand{\mmns}{\ensuremath{\text{mm}}\xspace}

\newcommand{\distauh}{\ensuremath{\tauh^\text{dis}}\xspace}
\newcommand{\mutauh}{\ensuremath{\PGm\tauh}\xspace}
\newcommand{\ctauz}{\ensuremath{c\tau_{0}}\xspace}
\newcommand{\jo}{\ensuremath{{\text{j1}}}\xspace}
\newcommand{\jt}{\ensuremath{{\text{j2}}}\xspace}
\newcommand{\ptjt}{\ensuremath{p_{\text{T,\ j2}}}\xspace}
\newcommand{\dxy}{\ensuremath{d_{xy}}\xspace}
\newcommand{\dxysig}{\ensuremath{\hat{d}_{xy}}\xspace}
\newcommand{\frw}{\ensuremath{f_{\text{W}}}\xspace}
\newcommand{\frdy}{\ensuremath{f_{\text{DY}}}\xspace}
\newcommand{\frbr}{\ensuremath{f_{\text{BR}}}\xspace}

\newcommand{\mstau}{\ensuremath{m_{\PSGt}}\xspace}

\newcommand{\dyjets}{\ensuremath{\text{DY+jets}}\xspace}
\newcommand{\wjets}{\ensuremath{{\PW}\text{+jets}}\xspace}

\newcommand{\distau}{\textsc{DisTau}\xspace}

\newcommand{\rtz}{\ensuremath{\text{T0}}\xspace}
\newcommand{\rto}{\ensuremath{\text{T1}}\xspace}
\newcommand{\rtt}{\ensuremath{\text{T2}}\xspace}

\newcommand{\brtz}{\ensuremath{\text{BRT0}}\xspace}
\newcommand{\brto}{\ensuremath{\text{BRT1}}\xspace}
\newcommand{\brtt}{\ensuremath{\text{BRT2}}\xspace}

\newcommand{\rtofz}{\ensuremath{\text{T10}}\xspace}
\newcommand{\rttfz}{\ensuremath{\text{T20}}\xspace}
\newcommand{\rttfo}{\ensuremath{\text{T21}}\xspace}

\newcommand{\mc}{\ensuremath{\text{MC}}\xspace}

\newlength\cmsTabSkip\setlength{\cmsTabSkip}{1ex}
\newlength\cmsTabSkipSmall\setlength{\cmsTabSkipSmall}{0.5ex}
\newlength\cmsTabSkipLarge\setlength{\cmsTabSkipLarge}{4ex}

\DeclareRobustCommand{\neutralinos}{{\HepParticle{\PSGc}{i}{0}}\Xspace}
\DeclareRobustCommand{\charginos}{{\HepParticle{\PSGc}{i}{\pm}}\Xspace}
\DeclareRobustCommand{\slepton}{\HepSusyParticle{\Pell}{}{}\Xspace}

\newcolumntype{L}[1]{>{\raggedright\let\newline\\\arraybackslash\hspace{0pt}}m{#1}}
\newcolumntype{C}[1]{>{\centering\let\newline\\\arraybackslash\hspace{0pt}}m{#1}}
\providecommand{\cmsTable}[1]{\resizebox{\textwidth}{!}{#1}}

\cmsNoteHeader{EXO-24-020}
\title{Search for the pair production of long-lived supersymmetric partners of the tau lepton in proton-proton collisions at \texorpdfstring{$\sqrt{s}=13\TeV$}{sqrt(s) = 13 TeV}}

\date{\today}

\abstract{
Gauge-mediated supersymmetry-breaking models provide a strong motivation to search for a supersymmetric partner of the tau lepton (stau) with a macroscopic lifetime. Long-lived stau decays produce tau leptons that are displaced from the primary proton-proton interaction vertex, leading to an unconventional signature. This paper presents a search for the direct production of long-lived staus decaying within the CMS tracker volume in proton-proton collisions at $\sqrt{s}=13\TeV$, performed for the first time with an identification algorithm based on a graph neural network dedicated to displaced tau leptons. The data sample, corresponding to an integrated luminosity of 138\fbinv, was recorded with the CMS experiment at the CERN LHC between 2016 and 2018. This search excludes, at 95\% confidence level, stau masses, $\mstau$, in the 126--260 (90--425)\GeV range for a proper decay length of 50\mm in the maximally mixed (mass-degenerate) scenario, while for $\mstau = 200\GeV$, stau proper decay lengths are excluded in the range 21--94 (6--333)\mm. These results improve the exclusion limits compared to previous searches, and extend the parameter space explored in the context of supersymmetry.
}

\hypersetup{
pdfauthor={CMS Collaboration},%
pdftitle={Search for pair production of long-lived supersymmetric partners of the tau lepton in proton-proton collisions at sqrt{s}=13 TeV},
pdfsubject={CMS},
pdfkeywords={CMS, EXO, SUSY, long-lived, tau, ML}}

\maketitle 

\section{Introduction}
\label{sec:introduction}

Supersymmetry (SUSY)~\cite{Ramond:1971gb,Golfand:1971iw,Neveu:1971rx,Volkov:1972jx,Wess:1973kz,Wess:1974tw,Fayet:1974pd,Nilles:1983ge} is an extension of the standard model (SM) of particle physics that offers potential solutions to several of its open questions, \eg, it can naturally account for the observed Higgs boson mass without resorting to fine-tuning of the bare mass~\cite{tHooft:1979rat,Witten:1981nf,Dine:1981za,Dimopoulos:1981au,Dimopoulos:1981zb,Kaul:1981hi,Aad:2012tfa,Chatrchyan:2012ufa,Chatrchyan:2013lba,Aad:2014aba,Khachatryan:2014jba,Aad:2015zhl}. In SUSY models, each SM particle is assigned a corresponding superpartner with the same quantum numbers, but differing in spin by half a unit. The contributions of SM particles and their superpartners to quantum corrections of the Higgs boson mass could partially cancel and thereby resolve the fine-tuning problem. In SUSY models with $R$-parity conservation~\cite{Farrar:1978xj}, the lightest SUSY particle (LSP) is stable~\cite{djoua,carena} and could be a viable dark-matter candidate~\cite{darkmatter}. 

The superpartners of the electroweak gauge and Higgs bosons, referred to as bino, wino, and Higgsino, mix to form neutral and charged mass eigenstates, called neutralinos (\neutralinos) and charginos (\charginos). The SUSY partners of the leptons are called sleptons (\slepton), with stau (\sTau) referring to the tau lepton superpartner. For the third generation, the mass eigenstates ($\PSGtDo$, $\PSGtDt$) can have different masses (by definition, $\PSGtDo$ is the lighter one) and are a mixture of the gauge eigenstates ($\PSGtR$, $\PSGtL$). This mixing can become large and depends on the Higgsino mass parameter, the ratio of the vacuum expectation values of the two Higgs fields ($\tan \beta$), and the third-generation $A$-parameter ($A_{\PGt}$)~\cite{ParticleDataGroup:2024cfk}.

In gauge-mediated SUSY breaking~\cite{Giudice:1998bp,Meade:2008wd,Evans:2016zau} models, the gravitino (\PXXSG) mass is proportional to the ratio of the SUSY-breaking scale $F$ to the Planck energy scale. 
The \PXXSG is thus very light and is the LSP.
In such models, the stau is a preferred candidate for the next-to-LSP, and $\sTau\to\PGt\PXXSG$ is the only available decay mode of the stau. 
The $\PSGt{-}\PGt{-}\PXXSG$ coupling is suppressed by a factor of $1/F$, which leads to longer lifetimes for the stau.
These scenarios can be investigated by searching for staus at proton colliders.

Searches for prompt \sTau decays have been performed at the CERN LEP collider~\cite{ALEPH:2001oot,DELPHI:2003uqw,L3:2003fyi,OPAL:2003nhx}, excluding at 95\% confidence level (CL) some scenarios with \sTau masses, $\mstau$, up to about 90\GeV and neutralino masses up to about 80\GeV.
The ATLAS and CMS Collaborations have searched for both promptly decaying and long-lived staus in proton-proton ($\Pp\Pp$) collisions at $\sqrt{s}=13\TeV$ at the CERN LHC~\cite{ATLAS:2019gti,CMS:2019eln,ATLAS:2020wjh,CMS:2022syk,CMS:2021kdm,ATLAS:2024fub,ATLAS:2025fdm}.
These searches excluded promptly decaying \sTau masses below 400\GeV for nearly massless neutralinos or gravitinos, assuming the production of mass-degenerate $\PSGtL$ and $\PSGtR$.
The limits on \sTau masses for the long-lived searches at $\sqrt{s}=13\TeV$ vary depending on the proper decay length (\ctauz) of the stau, but are of the same order of magnitude as in the prompt scenarios.
A recent search for displaced leptons by the ATLAS Collaboration~\cite{ATLAS:2024vnc} included data collected at $\sqrt{s}=13.6\TeV$ to exclude \sTau masses up to 440\GeV for $\ctauz=90\mm$.
In a scenario with maximal mixing between $\PSGtL$ and $\PSGtR$, $\mstau$ values between 150 and 220\GeV have been excluded for $\ctauz=0.1\mm$ at CMS with $\Pp\Pp$ collisions at $\sqrt{s}=13\TeV$~\cite{CMS:2022syk}.
Both ATLAS and CMS have also performed searches at $\sqrt{s}=8\TeV$~\cite{ATLAS:2014hep,ATLAS:2015eiz,CMS:2016gjw}.

The existing searches for long-lived staus that decay sufficiently within the tracker volume can be broadly divided into two categories.
The first category consists of generic searches for long-lived charged particles with final states containing electrons or muons from leptonic \PGt decays, which provide interpretations for long-lived staus. These searches profit from the clean signatures of the leptons.
The second category consists of searches with hadronically decaying tau leptons (\tauh), which target the dominant \PGt decay mode, and are based on reconstruction and identification algorithms optimized for \tauh candidates produced from prompt decays of particles~\cite{Sirunyan:CMS-TAU-16-003,CMS:2022prd}.
Such algorithms lose efficiency rapidly as the transverse distance of the decay vertex from the beam axis increases, falling by ${\approx}95\%$ at around 10\mm~\cite{CMS-DP-2024-053}, thereby reducing the sensitivity of a search for staus with higher \ctauz.

\begin{figure}[!htb]
\centering
\includegraphics[width=0.55\textwidth]{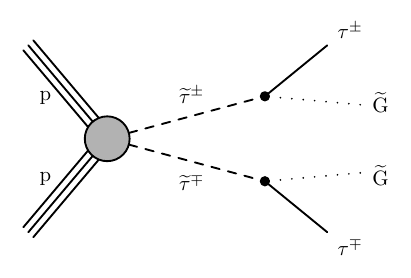}
\caption{Diagram of stau pair production in $\Pp\Pp$ collisions at the LHC, and the decay that leads to a final state with pairs of tau leptons accompanied by gravitinos.}
\label{fig:sig_diag}
\end{figure}

This paper focuses on the search for long-lived staus decaying into a \PGt and a nearly massless (1\GeV) \PXXSG, as shown in Fig.~\ref{fig:sig_diag}, within the framework of simplified models with masses between 90 and 600\GeV and \ctauz between 1 and 1000\mm~\cite{Alwall:2008ag,Alves:2011wf}.
Since tau leptons decay to hadrons more often than to electrons and muons~\cite{ParticleDataGroup:2024cfk}, this search targets the mode where both the tau leptons decay hadronically.
In order to overcome aforementioned limitations of existing \tauh identification algorithms, a dedicated algorithm based on the \textsc{ParticleNet}~\cite{PhysRevD.101.056019} graph convolutional neural network has been developed to identify \tauh produced far from the $\Pp\Pp$ collision vertex, \ie, displaced (\distauh).
This is referred to as the \distau identification algorithm~\cite{CMS-DP-2024-053}.

This paper is organized as follows. The CMS detector is introduced in Section~\ref{sec:cms}, and the event simulation is detailed in Section~\ref{sec:simulation}. Section~\ref{sec:reconstruction} contains the event reconstruction. The event selection and background estimation are described in Sections~\ref{sec:selection} and \ref{sec:background}, respectively. The systematic uncertainties are discussed in Section~\ref{sec:systematics}, and the results are presented in Section~\ref{sec:results}. A summary is given in Section~\ref{sec:summary}.
Tabulated results are provided in the HEPData record for this analysis~\cite{hepdata}.

\section{The CMS detector}
\label{sec:cms}

The central feature of the CMS apparatus is a superconducting solenoid of 6\unit{m} internal diameter that provides a magnetic field of 3.8\unit{T}. Within the solenoid volume are a silicon pixel and strip tracker, a lead tungstate crystal electromagnetic calorimeter (ECAL), and a brass and scintillator hadron calorimeter (HCAL), each composed of a barrel and two endcap sections. Forward calorimeters extend the coverage in pseudorapidity ($\eta$) provided by the barrel and endcap detectors. Muons are detected in gas-ionization chambers embedded in the steel flux-return yoke outside the solenoid.
A more detailed description of the CMS detector, together with a definition of the coordinate system used and the relevant kinematic variables, is provided in Refs.~\cite{Chatrchyan:2008zzk,CMS:2023gfb}.

Events of interest are selected using a two-tiered trigger system. The first level, composed of custom hardware processors, uses information from the calorimeters and muon detectors to select events at a rate of around 100\unit{kHz} within a fixed latency of 4\mus~\cite{CMS:2020cmk}. The second level, known as the high-level trigger, consists of a farm of processors running a version of the full event reconstruction software optimized for fast processing, and it reduces the event rate to a few\unit{kHz} before data storage~\cite{CMS:2016ngn,CMS:2024aqx}.

\section{Event simulation}
\label{sec:simulation}
Several Monte Carlo (MC) event generators are used to simulate events for the signal and background processes. The signal process shown in Fig.~\ref{fig:sig_diag} is simulated based on simplified SUSY models.
The simulation is performed at leading order (LO) using the \MGvATNLO v2.4.2 event generator~\cite{Alwall:2014hca}.
The NNPDF3.0 parton distribution function (PDF) set~\cite{NNPDF:2014otw} is used for simulated samples corresponding to the 2016 data-taking period, whereas NNPDF3.1~\cite{NNPDF:2017mvq} is used for 2017 and 2018.
The signal cross sections are evaluated for a set of discrete $\mstau$ values between 90 and 600\GeV at next-to-LO (NLO) in perturbative quantum-chromodynamics calculations (pQCD), using next-to-leading logarithmic (NLL) soft-gluon threshold resummation, as provided by \textsc{Resummino}~\cite{Fuks:2013lya, Fuks:2013vua, Fiaschi:2023tkq, Fiaschi:2018xdm, Bozzi:2007tea, Bozzi:2006fw}.
The cross sections for $\mstau$ between the values that are reported in the aforementioned references are interpolated using the \textsc{HEPi-fast} tool~\cite{Neuwirth2024-vc}.
For each $\mstau$ hypothesis, signal samples have been simulated with stau \ctauz values ranging from 1 to 600\mm.
Two different cases have been considered for the stau pair production in this study.
The first is the production of $\PSGtDo\HepSusyAntiParticle{\PGt}{1}{}$ in a ``maximally mixed scenario", where the mixing angle between $\PSGtL$ and $\PSGtR$ is $\pi/4$, and $\PSGtDt$ is considered to be too massive to be produced at the LHC.
The second scenario is the production of both $\PSGtL\HepSusyAntiParticle{\PGt}{L}{}$ and $\PSGtR\HepSusyAntiParticle{\PGt}{R}{}$ with no mixing between the gauge eigenstates.
Since in this case the gauge eigenstates are the same as the mass eigenstates, which are considered to be degenerate in mass, this is referred to as the ``mass-degenerate scenario" in this paper.
The mass-degenerate scenario leads to a cross section between three and five times higher than the maximally mixed scenario for each $\mstau$ considered.
This is because in the s-channel production of stau pairs via $\PZ/\PGg^{*}$, $\PSGtL$ has a stronger coupling to the \PZ boson than $\PSGtR$, but both couple equally to the photon.

In this search, the contribution from background processes is estimated using control samples in data, as detailed in Section~\ref{sec:background}.
Simulated events are used to validate the background estimation procedure based on data, and to train and calibrate the identification algorithm discussed in Section~\ref{sec:reco_distau}.
Simulated samples of top quark-antiquark pair production (\ttbar), single top quark production, and {\ttbar}\PV (\PV = \PW or \PZ) events are generated with the \POWHEG 2.0 generator~\cite{Oleari:2010nx, Nason:2004rx, Frixione:2007vw, Alioli:2010xd, Frixione:2007nw} at NLO in pQCD.
Diboson ($\PV\PV$ and $\PV\PH$) events are simulated at NLO with the \MGvATNLO 2.6.5 generator. The \MADGRAPH (v2.2.2 for 2016, v2.4.2 for 2017 and 2018)~\cite{Alwall:2014hca} event generator is used at LO for modeling the Drell--Yan+jets (\dyjets) and \wjets processes, as well as for SM events comprised uniquely of jets produced through the strong interaction, referred to as multijet events.

The kinematic variables are generated including representative distributions of additional $\Pp\Pp$ interactions per bunch crossing, referred to as pileup.
These samples are produced separately for each year of data taking to account for different pileup and detector conditions in the three years.
Additionally, the simulated samples for each year are reweighted such that their pileup profiles match the one measured in the data of the corresponding year.

The CMS detector response for all processes is modeled using \GEANTfour~\cite{GEANT4:2002zbu}, and simulated events are then reconstructed in the same way as collision data.
For all simulated events, the description of parton showering and hadronization, as well as the hadronic decays of tau leptons, is performed with \PYTHIA v8.212~\cite{Sjostrand:2014zea} with the CP5 underlying event tune~\cite{CMS:2019csb}.

Small discrepancies observed between simulation and data are removed by using several correction factors, as discussed in Section~\ref{sec:systematics}.

\section{Event reconstruction}
\label{sec:reconstruction}

\subsection{Standard object reconstruction}
\label{sec:reco_obj}

A particle-flow (PF) algorithm~\cite{CMS-PRF-14-001} reconstructs each individual particle in an event, with an optimized combination of information from various components of the CMS detector.
The energies of electrons and photons are measured in the ECAL.
The momentum of electrons is determined by a combined measurement of the track momentum in the tracker, the energy of the matching ECAL supercluster, and the energy of all bremsstrahlung photons consistent with originating from the track.
The momentum of muons is obtained from the bending of the corresponding tracks in the tracker and the muon spectrometer, which comprises three technologies: drift tubes, cathode strip chambers, and resistive plate chambers.
The energy of charged hadrons, which is corrected for the response function of the calorimeters to hadronic showers, is determined from a combination of the momentum measured in the tracker and the energy of the matching ECAL and HCAL clusters.
Finally, the energy of neutral hadrons is obtained from the corrected energies in the corresponding ECAL and HCAL clusters.

Collision vertices are reconstructed by fitting tracks in an event and are required to be within 24\cm of the center of the detector in the $z$ direction (along the beam) and to have a transverse displacement from the beam line of less than 2\cm.
The primary vertex is taken to be the vertex corresponding to the hardest scattering in the event, evaluated using tracking information alone, as described in Section 9.4.1 of Ref.~\cite{CMS-TDR-15-02}.
Reconstruction of jets is performed by clustering PF objects using the anti-\kt algorithm~\cite{Cacciari:2008gp,Cacciari:2011ma} with a distance parameter of $R=0.4$.
The jet momentum is determined as the vector sum of all particle momenta in the jet and is typically within 5--10\% of the true momentum over the entire \pt spectrum within the detector acceptance.
Pileup interactions contribute to spurious tracks and calorimetric energy deposits, increasing the apparent jet momentum.
To mitigate this effect, tracks identified as originating from pileup vertices are discarded, and an offset is applied to correct for the remaining contributions~\cite{Khachatryan:2016kdb}.
Jets are calibrated using information from both simulation and data~\cite{Khachatryan:2016kdb}.
Additional selection criteria are applied to remove jets that are potentially dominated by instrumental effects or reconstruction failures~\cite{CMS-PAS-JME-16-003}.
Only jets with $\pt>30\GeV$, $\abs{\eta}<2.1$, and containing at least one charged PF object with $\pt>1\GeV$ are considered in this analysis.
The absolute transverse impact parameter ($\abs{\dxy}$) of a jet and its impact parameter significance ($\dxysig$)---the ratio between $\abs{\dxy}$ and its uncertainty---are defined as the $\abs{\dxy}$ and $\dxysig$ of its charged constituent with the highest \pt.
For nonisolated particles with $1 < \pt < 10\GeV$ and $\abs{\eta} < 3.0$, the track resolutions are typically 1.5\% in \pt and 20--75\mum in \dxy~\cite{DP-2020-049}.
Isolated particles with \pt = 100\GeV and $\abs{\eta} < 1.4$ have track resolutions of 2.8\% in \pt and 10\mum in \dxy~\cite{CMS:2014pgm}.
Choosing the charged constituent with the highest \pt in jets originating from the fragmentation of partons may not be robust against infrared and collinear emissions in the parton shower simulation.
However, this is expected to have a negligible impact on the analysis as the contribution of such jets to the background is estimated entirely using data, as detailed in Section~\ref{sec:background}.

Jets originating from the fragmentation of \PQb quarks are identified as \PQb-tagged jets~\cite{CMS:2017wtu} using the \textsc{DeepJet} algorithm~\cite{CMS-DP-2023-005, Bols:2020bkb}. The algorithm employs properties of reconstructed secondary vertices and charged and neutral particle constituents of the jet as inputs to a convolutional deep neural network.
The ``medium" selection or working point (WP) of this algorithm corresponds to a signal efficiency of about 80\%, with a misidentification probability of about 1\% for light-flavored jets (\ie, originating from gluons or up, down, and strange quarks) and about 11\% for jets originating from charm quarks.
Jets passing this WP are henceforth referred to as \PQb-tagged jets.

Electrons are identified using the ``loose" WP of a boosted decision tree algorithm~\cite{CMS:2020uim,Khachatryan:2015hwa} that uses inputs based on the spatial distribution of the shower, track-to-cluster-matching criteria, consistency between the cluster energy and the track momentum, and isolation.
Electrons satisfying the loose WP and with $\pt > 20\GeV$ and $\abs{\eta} < 2.4$ are referred to as veto-electrons in what follows.

Muon reconstruction uses a global fit combining information from the tracker and muon spectrometers~\cite{Sirunyan:2018_1804.04528}.
Muon candidates are required to pass either the ``loose" or the ``tight" WP of an identification algorithm using criteria such as the quality of the global fit and the geometrical matching between the tracks in the tracker and the muon spectrometers.
Isolation criteria are imposed on the muon candidates to reject those originating from hadronic decays.
The isolation variable used for this purpose is defined as the scalar \pt sum of reconstructed charged and neutral particles within a cone of radius $ \Delta R = \sqrt{\smash[b]{(\Delta\eta)^2+(\Delta\phi)^2}}$ = 0.4, where $\phi$ is the azimuthal angle in radians, around the muon candidate track, which is excluded from the sum, divided by the \pt of the lepton candidate.
Charged particles not originating from the primary vertex are excluded from this sum, and a correction is applied to account for the neutral components originating from pileup, as described in Ref.~\cite{CMS:2020ebo}.
Muons satisfying the tight (loose) WP and with relative isolation less than 15 (25)\%, $\pt > 20\ (28)\GeV$, and $\abs{\eta} < 2.4$, are referred to as veto-muons (id-muons) in what follows.

Hadronic tau lepton candidates are reconstructed from one charged hadron and up to two neutral pions, or three charged hadrons and up to one neutral pion, consistent with originating from the decay of a tau lepton, using the hadrons-plus-strips algorithm~\cite{Sirunyan:CMS-TAU-16-003}.
To distinguish between jets originating from quarks or gluons and genuine hadronic tau lepton decays, the discriminant of the \textsc{DeepTau}~\cite{CMS:2022prd} deep neural network algorithm is used.
The \tauh candidates are required to have $\pt>30\GeV$, $\abs{\eta}<2.4$, and are selected with the ``medium" WP of the above discriminant, which has an efficiency of ${\approx}$70\% and a misidentification probability of ${\approx}$1\%.
In addition, the \tauh candidates are required to satisfy the ``very loose" (``medium") WP of the \textsc{DeepTau} discriminant against electrons (muons), which has an efficiency of ${\approx}$98 (99.8)\% and a misidentification probability of ${\approx}$1.5 (0.04)\%.
These \tauh candidates are used to identify a region enriched in $\PZ\to\PGt\PGt$ events, where the calibration factors for the \distau identification efficiency are measured, as described in Section~\ref{sec:reco_distau}.

The missing transverse momentum vector, \ptvecmiss, is computed as the negative vector \ptvec sum of all the PF objects in an event and its magnitude is denoted as \ptmiss~\cite{Sirunyan:2019kia}.
The \ptvecmiss is modified to account for the energy calibration of all the PF candidates in an event, clustered into jets or not.

\subsection{Displaced \texorpdfstring{\tauh}{tauh} candidates}
\label{sec:reco_distau}

The \distau~\cite{CMS-DP-2024-053} graph neural network algorithm is trained to distinguish between signal jets originating from displaced \tauh candidates (\distauh) and those from the fragmentation of light-flavored, charm, and bottom quarks, and gluons (background jets).
The signal jets used for the training are selected from three different signal samples corresponding to \ctauz values of 1, 100, and 1000\mm.
These samples are orthogonal to the ones used for obtaining the signal yields in Section~\ref{sec:results}.
To use well-reconstructed signal jets for the training, the longitudinal (transverse) position of the stau decay vertex is required to be within 1000\mm (500\mm) of the interaction point, which is approximately the longitudinal (radial) extent of the inner tracker detector.
No lower threshold is imposed on the signal jet displacement.
Hence, the network is allowed to learn from signal jets with a wide range of displacements, arising from staus that decay within the inner tracker volume.
The background jets are taken from multijet events, and jets with different flavors are treated together as a single background category, as separating them by flavor did not provide any significant improvement in discrimination.
Jets from \tauh candidates produced from prompt decays of particles are not treated as a separate background category as there is no distinctive difference in properties between them and signal jets with very small displacements.
Moreover, the contribution from prompt \tauh produced in SM processes being misidentified as \distauh candidates is sufficiently suppressed by the selection on jet \dxy, as discussed in Sections~\ref{sec:selection} and \ref{sec:background}.

The algorithm, based on the \textsc{ParticleNet}~\cite{PhysRevD.101.056019} architecture, represents the particles in a jet as a ``point cloud'' and performs dynamic graph convolution operations~\cite{10.1145/3326362} on the properties of each jet constituent and its nearest neighbors.
The features provided as input to the network are the same as those used in the \textsc{DeepTau} algorithm~\cite{CMS:2022prd}.
The \distau algorithm is optimized to identify \distauh produced within the tracker volume of the CMS detector.
The algorithm exploits features relevant for both hadronic \PGt decays (such as the number of charged hadrons and their isolation properties), as well as displaced \PGt decays (such as the impact parameters of the charged constituent tracks).
Impact parameters are also used by the \textsc{DeepTau} algorithm as input features, but they play a more significant role in the discriminating power of the \distau algorithm.
Displacements of less than 10\mm result in reduced discrimination between \distauh candidates and background jets, while for displacements above around 500\mm, the \distauh candidate cannot produce enough hits in the tracker to be distinguished from the track of the \sTau that produced it, leading to a lower identification efficiency.

\begin{figure}[!htb]
    \centering
    \includegraphics[width=0.7\linewidth]{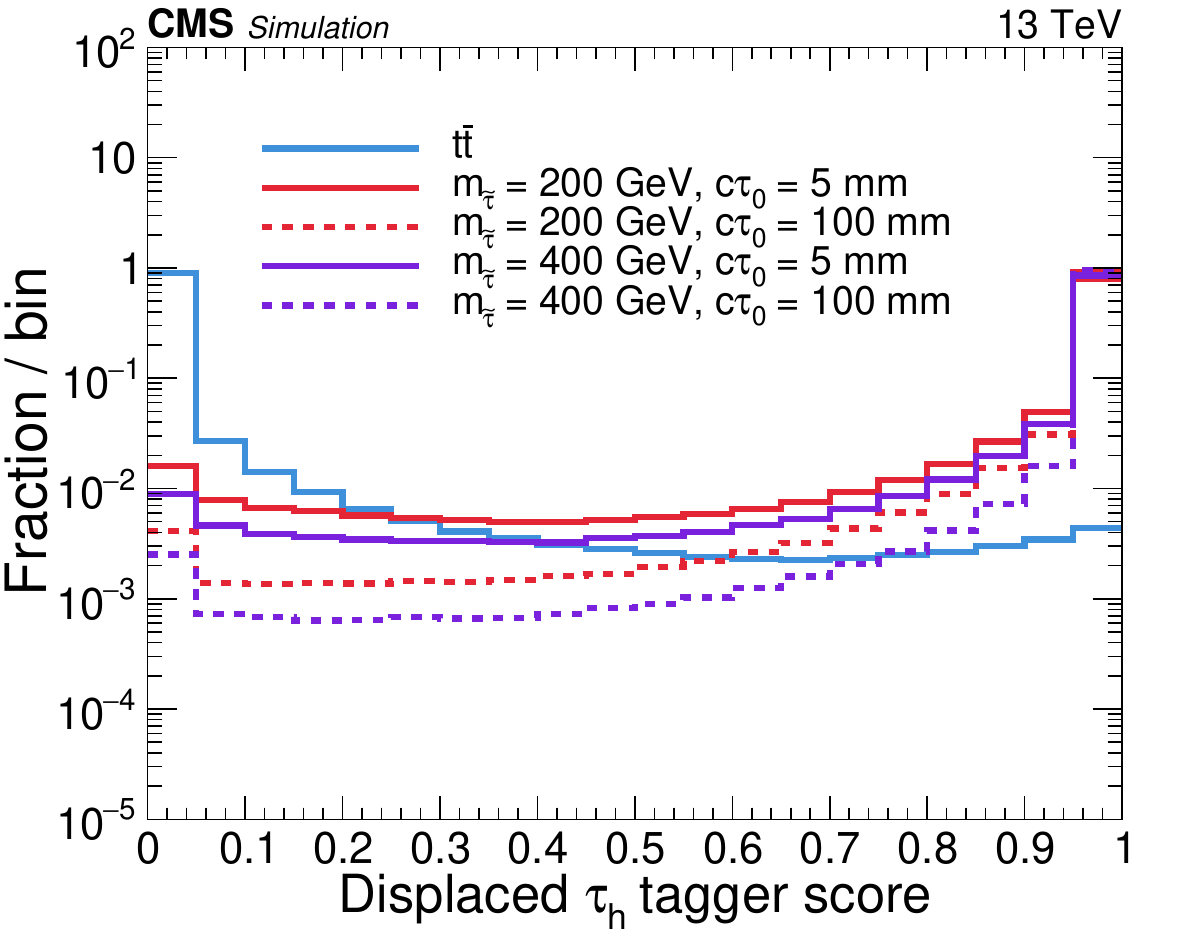}
    \caption{
    Distributions of \distau score for signal and background jets.
    The background jets are taken from simulated \ttbar events where both top quarks decay hadronically, while signal jets are sampled from four representative $(\mstau~[{\GeVns}],\ \ctauz~[{\mmns}])$ hypotheses: (200, 5), (200, 100), (400, 5), and (400, 100).
    Each distribution is normalized such that its integral is unity.
    }
    \label{fig:distau_score_mc}
\end{figure}

The distributions of \distau scores for signal and background jets are presented in Fig.~\ref{fig:distau_score_mc}.	
The ``loose", ``medium", and ``tight" WPs of the \distau algorithm are denoted as WPL, WPM, and WPT, with signal efficiencies of about ${>}99$, 98, and 80\%, respectively, measured in simulated events for a signal hypothesis of $\mstau=200\GeV$ and $\ctauz=100\mm$.
Their respective misidentification probabilities, evaluated in simulated \ttbar events, are approximately 7, 1, and 0.06\%.
The efficiencies (misidentification probabilities) quoted here are inclusive, \ie, measured with respect to all signal (background) jets in the simulated sample.
The lower thresholds on the tagger score corresponding to the three WPs are 0.05, 0.7, and 0.99, respectively.
To be chosen as a \distauh candidate, a jet selected with the criteria mentioned in Section \ref{sec:reco_obj} is also required to satisfy $\abs{\dxy}>2\mm$, $\dxysig>5$, and pass the WPL selection.
The requirement on $\dxysig$ ensures that the $\dxy$ of the selected \distauh candidates is well measured.

The efficiency of the \distau identification algorithm for the simulated signal is corrected to match the efficiency measured in data, using a control region (CR) enriched in DY($\PZ/\PGg^{*}\to\PGt\PGt$)+jets events, where one of the tau leptons decays to a muon and the other decays hadronically.
This is referred to as the \mutauh CR.
As there are no truly signal-like \distauh candidates in SM processes, the \tauh in the event serves as a proxy to evaluate the identification efficiency.
Following a strategy similar to that in Ref.~\cite{CMS:2017zyp}, events are selected online with a single-muon trigger, and required to have exactly one id-muon and one \tauh candidate with opposite electric charges, angular separation $\Delta R(\PGm, \tauh) > 0.5$, invariant mass of the \mutauh system between 60 and 80\GeV, transverse mass $\mT(\PGm, \ptvecmiss) < 60\GeV$, and $D_{\zeta}(\PGm, \tauh, \ptvecmiss) > -25\GeV$, with  $D_{\zeta}(\PGm, \tauh, \ptvecmiss)$ defined as~\cite{CMS:2014ccx, CMS:2017zyp}

\begin{linenomath}
\begin{equation}
D_{\zeta}(\PGm, \tauh, \ptvecmiss) = \ptvecmiss\cdot\hat{\zeta} - 0.85 [\ptvec(\PGm)+\ptvec(\tauh)] \cdot\hat{\zeta} , 
\label{eq:dzeta}
\end{equation}
\end{linenomath}
where $\hat{\zeta}$ is the unitary vector oriented along the bisector of the muon and the \tauh candidate transverse momenta, $\ptvec(\PGm)$ and $\ptvec(\tauh)$, respectively.
The transverse mass for two objects labeled ``a" and ``b" is defined as
\begin{linenomath}
\begin{equation}
\mT^{2}(\text{a},\text{b}) =
m^{2}_\text{a} + m^{2}_\text{b} + 2(\et^\text{a} \et^\text{ b} - \ptvec^\text{ a} \cdot \ptvec^\text{ b}) . 
\label{eq:mT}
\end{equation}
\end{linenomath}
Here the masses (transverse momenta) of the particles are denoted by $ m_{\text{a}} $ and $ m_{\text{b}} $ ($ \ptvec^{\text{ a}} $ and $ \ptvec^{\text{ b}} $), respectively, and $\et^{2} = m^{2} + \pt^{2}$.
The $\mT(\PGm, \ptvecmiss)$ formulation is then obtained by replacing the generic particles with a muon and a massless particle of transverse momentum \ptvecmiss.

\begin{figure}[!htb]
    \centering
    \includegraphics[width=0.7\linewidth]{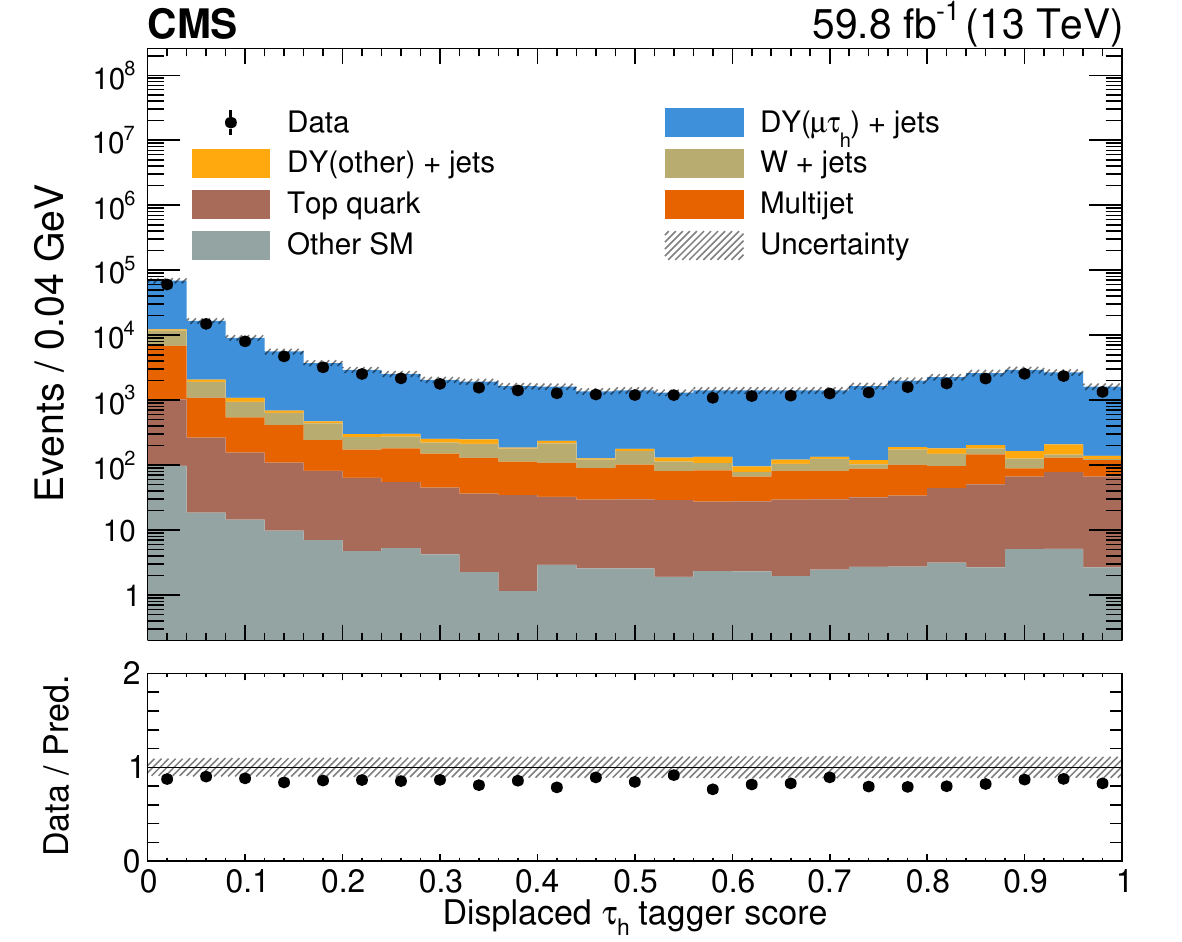}
    \caption{
    Distributions of \distau score for \distauh probes in the \mutauh CR described in Section~\ref{sec:reco_distau}, for data and predicted SM processes, corresponding to the 2018 data-taking period.
    Here DY(\mutauh) represents events from the $\PZ/\PGg^{*}\to\PGt\PGt$ process, where one of the tau leptons decays to a muon and the other decays hadronically.
    Events from other decay modes of $\PZ/\PGg^{*}$ are denoted as DY(other).
    Processes denoted as ``Top quark'' comprise \ttbar, single top, and {\ttbar}V, and other SM processes include events from diboson production.
    Uncertainties in the simulation are shown as a grey band and include only the statistical uncertainty in the number of simulated events.
    Simulation is not fully calibrated to describe data, as the distribution is shown to illustrate the general behavior of the classifier and is not directly used for the measurement of simulation-to-data correction factors.
    The bars representing the statistical uncertainty in the data are not visible as they are smaller than the marker size.
    The results for the 2016 and 2017 data-taking periods are similar in behavior.
    }
    \label{fig:distau_score_mutauh}
\end{figure}

Correction factors are estimated using the ``tag-and-probe'' technique described in Refs.~\cite{Sirunyan:CMS-TAU-16-003,CMS:2010svw, CMS:2022prd}.
The muon serves as the ``tag'' in this procedure, while the \distauh candidate nearest to the \tauh and satisfying $\Delta R(\tauh, \distauh) < 0.4$ is treated as the ``probe''.
The distribution of the \distau score for the probes in the \mutauh CR is shown in Fig.~\ref{fig:distau_score_mutauh}.
To measure the simulation-to-data correction for a given WP, a maximum likelihood fit is performed on the invariant mass distribution of the \mutauh system, which offers a bounded (60--80\GeV, as stated above) distribution that includes probes of any momenta, provided they satisfy the requirements on jet \pt and $\abs{\eta}$ discussed in Section~\ref{sec:reco_obj}.
The fit is performed simultaneously for events in two regions: one in which the probe passes the chosen WP and one in which it fails.
The correction to the \distau efficiency is then treated as an anticorrelated parameter across the two regions in the fit, in which the efficiency of the selected WP in simulation is corrected to match the one in data.
This procedure yields a correction factor that can be applied to simulation for each \distauh candidate (satisfying the chosen WP) in an event.
The correction factor per \distauh candidate for WPT is evaluated from the fit to be within 10\% of unity for all years with an uncertainty of 0.15.
The leading contribution to this uncertainty arises from the statistical uncertainty in the number of events that pass the WPT selection.
The behavior of the correction factors with increasing probe jet \dxy has also been studied, and no significant dependence is observed.
In Fig.~\ref{fig:distau_score_mutauh}, the small departures from unity in the ratio between data and predicted SM processes (${\approx}$10--15\% in most bins) do not affect the measurement of the correction factors to any appreciable degree, as the fit is intended to measure the difference in the efficiency of the WPT selection between data and simulation, and not the overall normalization difference.

\section{Event selection}
\label{sec:selection}

In order to reject the vast majority of events that are easily identifiable as incompatible with the signal topology, we define a baseline region (BR) containing events in both the signal-enriched search region (SR) and the sideband regions that are used for estimating the background contribution in the SR.

Since the signal is expected to have high missing momentum from gravitinos that escape without interacting with the detector, events in the BR are selected with high-level triggers requiring a large transverse momentum imbalance: $\ptmiss>100\GeV$ in 2016 and $\ptmiss>120\GeV$ in 2017 and 2018.
The efficiency of the \ptmiss triggers increases from about 70\% at 120\GeV to a plateau of almost unity at 250\GeV.
Offline, events are required to have $\ptmiss > 120\GeV$ and exactly two \distauh candidates.
Since the offline requirement on \ptmiss is in the turn-on region of the trigger efficiency, the uncertainty in the trigger efficiency is treated differently for events with $\ptmiss$ below and above 250\GeV, as described in Section~\ref{sec:systematics}. 
Events are rejected if they have any veto-electron, veto-muon, or \PQb-tagged jet.
The leading (higher-\pt) and subleading (lower-\pt) \distauh candidate jets in an event are referred to as \jo and \jt, respectively.
The background contribution from jets produced in noncollision processes, such as beam halo interactions, is greatly mitigated by requiring events to satisfy the beam halo filters~\cite{Sirunyan:2019kia}.
Since beam halo jets usually populate the plane defined by the beam axis and $\phi=0$ or $\pi$, their contribution is further reduced by requiring events to satisfy $\Delta\phi(\jo, \jt)>0.5$.
In order to mitigate the impact of increased noise in the ECAL endcaps observed during the 2017 data-taking period, events with $\HT^{\text{soft, fwd}} > 50\GeV$ are vetoed.
Here $\HT^{\text{soft, fwd}}$ is defined as the scalar sum of the \pt of all soft (uncorrected $\pt < 50\GeV$) jets in the forward region ($2.65 < \abs{\eta} < 3.139$).
Since jets in signal processes are typically harder and in the central region, this selection has a negligible effect on signal events, while suppressing background events.
The requirement on $\HT^{\text{soft, fwd}}$ thus helps to improve sensitivity to the signal, and is applied to all data-taking periods.

A separate CR enriched in \wjets events (WCR) is selected using single-muon triggers.
Offline, events are required to satisfy the aforementioned beam halo filters, have exactly one id-muon, no additional veto-muons or veto-electrons, $\ptmiss > 50\GeV$, $60 < \mT(\PGm, \ptvecmiss) < 120\GeV$, and exactly two \distauh candidates satisfying the same requirement on $\Delta\phi(\jo, \jt)$ as in the BR.
The region thus defined is about 85\% pure in \wjets events.

Single-muon triggers are also used to identify another separate CR enriched in \dyjets (DYCR).
Offline, events are required to satisfy the aforementioned beam halo filters, have exactly two id-muons with opposite charges, no additional veto-muons or veto-electrons, dimuon invariant mass between 70 and 120\GeV, and the same requirement on the \distauh candidates as the WCR.
This region is found to be approximately 96\% pure in \dyjets events.

The contamination from signal processes in WCR and DYCR is found to be negligible (${<}1\%$) using simulation. 
Events satisfying the BR, WCR, and DYCR selections are categorized into the T0, T1, and T2 regions based on whether zero, one, or two jets, respectively, passed the \distau WPT, as illustrated in Table~\ref{tab:region_defs}.
For example, \brtz includes events in the BR satisfying the T0 selection.
We use \brtt as the SR, whereas \brtz, \brto, WCR, and DYCR are CRs used for background estimation and validation, as described in Section~\ref{sec:background}.
All the regions discussed here are mutually orthogonal.

\begin{table}[!htbp]
\topcaption{Signal and control region definitions. The observed number of events in each region is shown in parentheses.}
\renewcommand\arraystretch{1.3}
\cmsTable{\begin{tabular}{l c c c}
Tag criteria & \rtz & \rto & \rtt \\[\cmsTabSkip]
Description
& \centerstack{Exactly two \distauh candidates\\ of which none pass WPT}
& \centerstack{ Exactly two \distauh candidates of\\  which only one passes WPT}
& \centerstack{ Exactly two \distauh candidates\\ of which both pass WPT} \\[\cmsTabSkip]
\hline
Baseline selections & \brtz (6287) & \brto (1384) & \brtt $ \equiv $ SR (81) \\
{\PW}+jets CR selections & WCR\rtz (5517) & WCR\rto (1094) & WCR\rtt (86) \\
DY+jets CR selections & DYCR\rtz (1156) & DYCR\rto (218) & DYCR\rtt (18) \\
\hline
\end{tabular}}
\label{tab:region_defs}
\end{table}

The dominant sources of \ptmiss in the signal events would be the weakly interacting gravitinos, whose kinematic properties are correlated with those of the visible objects, namely, the \tauh candidates.
In contrast, \ptmiss in the SM background processes primarily originates from neutrinos.
This difference can be exploited by first constructing the transverse mass $\mT(\text{vis}, \text{inv})$, defined in Eq.~\ref{eq:mT}, with the visible (``vis") and the invisible (``inv") particles. 
The distribution of $\mT(\text{vis}, \text{inv})$ presents a maximum at the mass of the parent of the visible and the invisible particles when there is only one source of missing momentum in the system.
To account for pair-produced particles where both have visible and invisible decay products, the ``stransverse mass (\mTii)"~\cite{Lester:1999tx, Barr:2003rg} is defined as:
\begin{linenomath}
\begin{equation}
\mTii^{2} (\text{vis1}, \text{vis2}, \ptvecmiss) = \min_{\ptvec^\text{ inv1} + \ptvec^\text{ inv2} = \ptvecmiss} [\max \{ \mT^{2}(\text{vis1},\text{inv1}), \mT^{2}(\text{vis2},\text{inv2}) \}] .
\label{eq:mT2}
\end{equation}
\end{linenomath}
Since the momenta of the individual invisible particles in Eq.~\eqref{eq:mT2} are unknown, \ptvecmiss is divided into two components ($ \ptvec^\text{ inv1} $ and $\ptvec^\text{ inv2} $) in such a way that the value of \mTii is minimized.
If \mTii is computed using the two $\distauh$ candidates as the visible objects, ``vis1'' and ``vis2'', then its upper limit in the signal will be the stau mass.
This is different from the SM background processes.
For example, in $\ttbar\to\PQb\PAQb\PWp\PWm$ events, where both the \PW bosons decay to a lepton (visible) and a neutrino (invisible), the upper limit is at the \PW boson mass.
In searches where the masses of the invisible particles are unknown, the calculation of \mTii requires an assumption on their masses~\cite{Barr:2009wu}.
In this analysis, we consider the gravitino to be massless when calculating \mTii.

\begin{figure}[h!]
    \centering
    \includegraphics[width=0.49\linewidth]{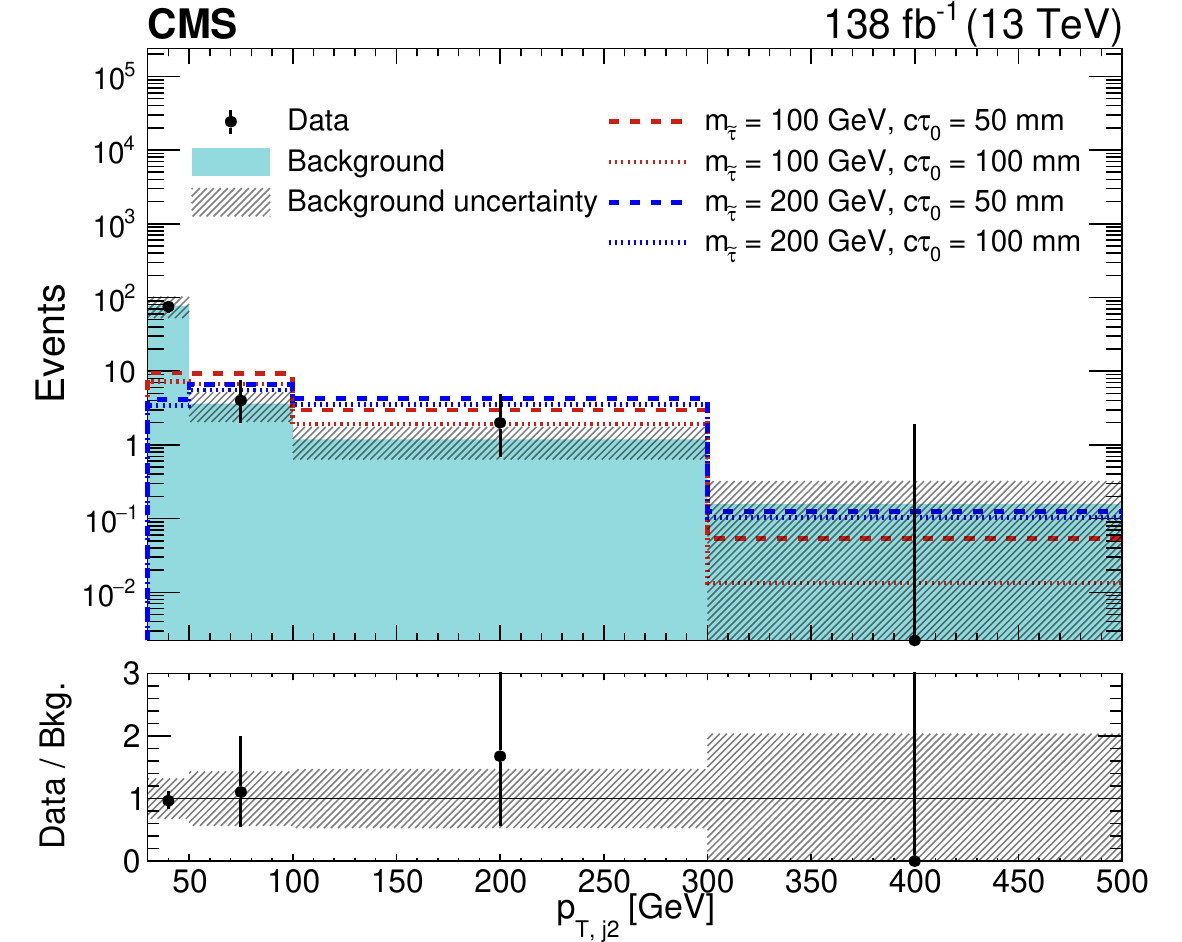}
    \includegraphics[width=0.49\linewidth]{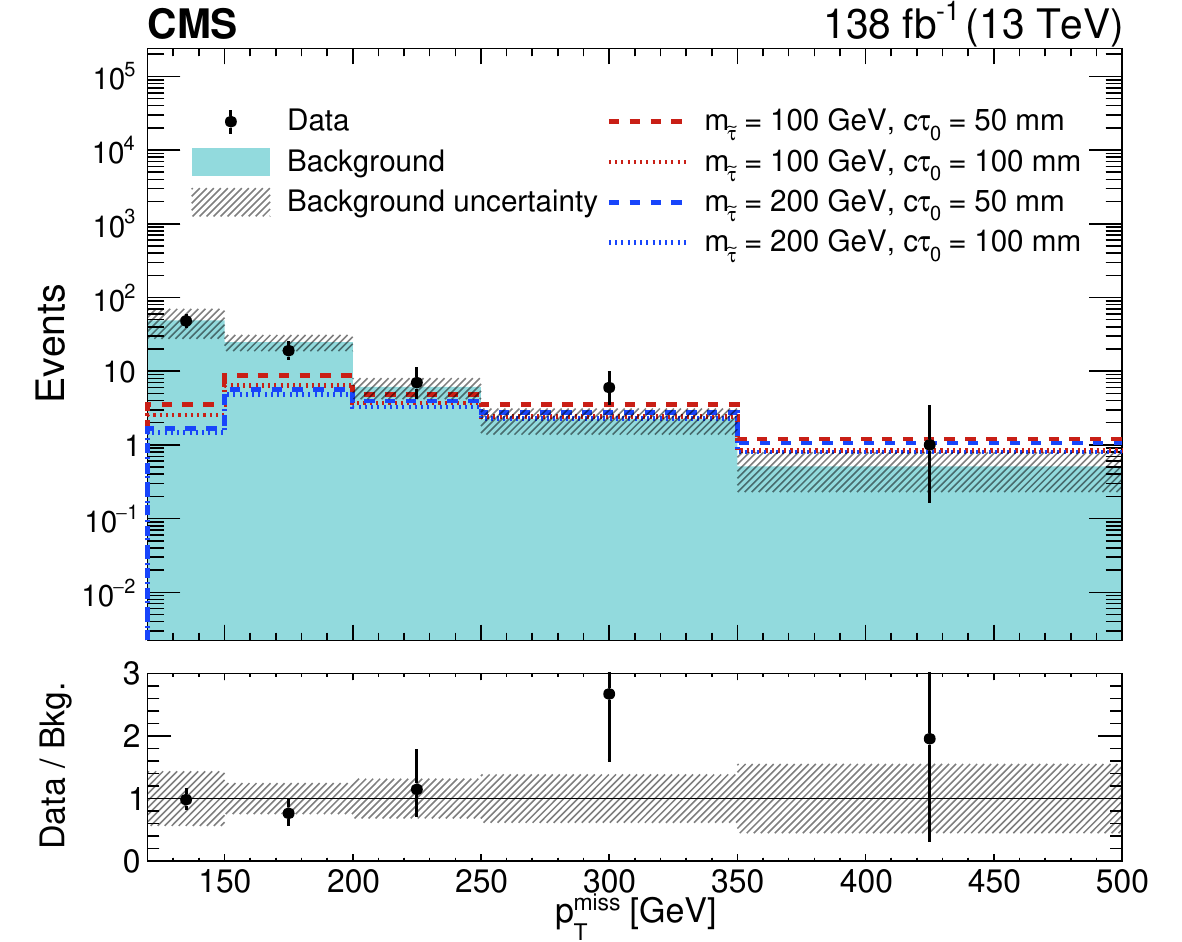}
    \\
    \includegraphics[width=0.49\linewidth]{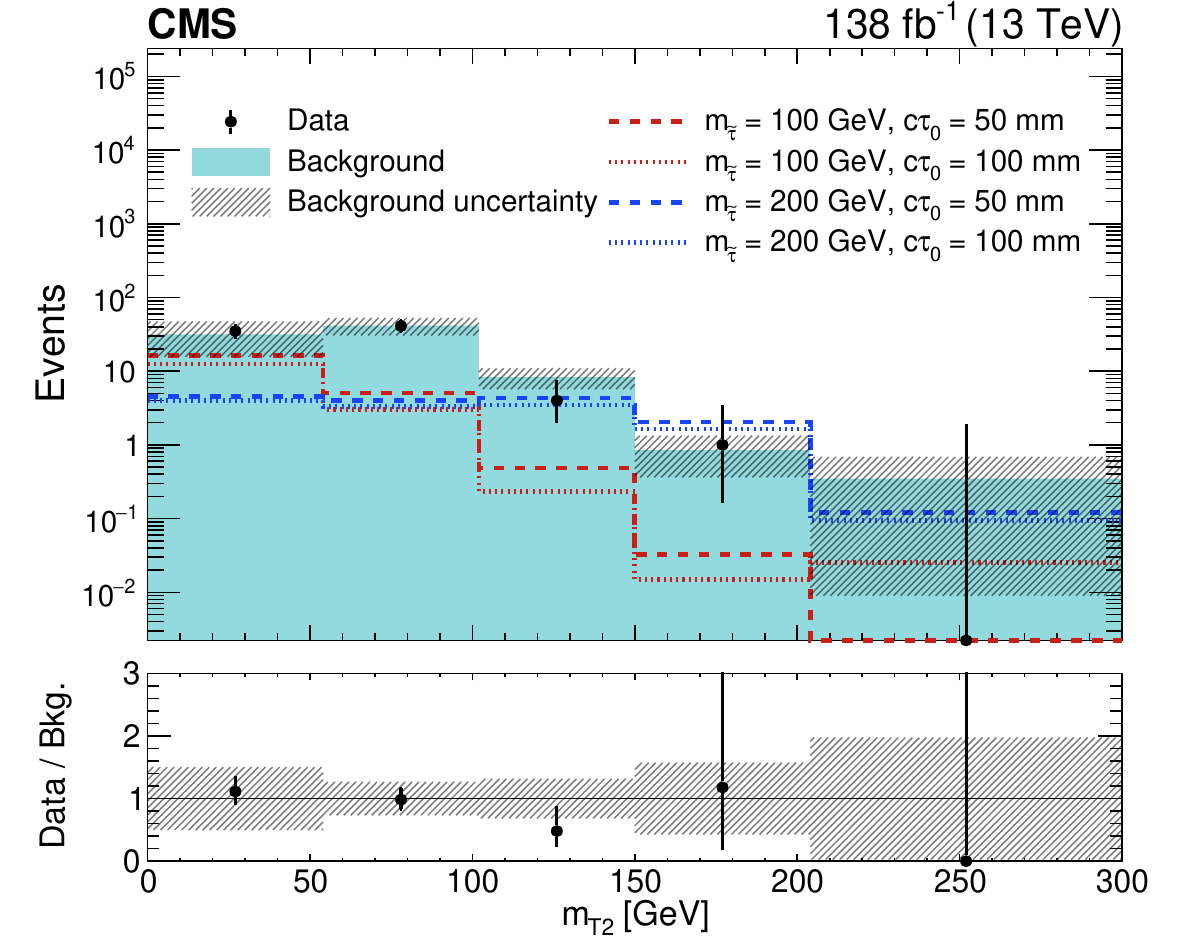}
    \caption{Distributions of the variables used to define the signal region for data and the predicted background: \ptjt (upper left), \ptmiss (upper right), and \mTii (lower).
    The signal distributions expected in the maximally mixed scenario for four representative sets of $(\mstau~[{\GeVns}],\ \ctauz~[{\mmns}])$ values are overlaid: (100, 50), (100, 100), (200, 50), and (200, 100).
    In bins where the observed yield is zero, the Garwood interval at 68\% \CL is shown as a positive uncertainty.
    The lower panel indicates the ratio of the observed number of events to the total predicted number of background events.
    The shaded bands indicate the statistical and systematic uncertainties in the predicted backgrounds, added in quadrature.
    The last bin includes the overflow.}
    \label{fig:distributions}
\end{figure}

In addition to \ptmiss and \mTii, the \pt of the subleading \distauh candidate (\ptjt) is found to be useful to discriminate between signal and background.
Distributions of these three variables for events within the SR are presented in Fig.~\ref{fig:distributions}.
The background distributions have been estimated using the procedure described in Section~\ref{sec:background}.
Signal events with different $\mstau$ populate different regions of the parameter space.
For example, regions with low (high) \ptmiss, \mTii, and \ptjt are sensitive to signals with low (high) $\mstau$.
In order to have high sensitivity over a wide range of $\mstau$, selected events are categorized into eight bins in the measured \ptmiss, \mTii, and \ptjt, as shown in Table~\ref{tab:bins}.

\begin{table}[!htbp]
\topcaption{The \ptjt, \ptmiss, and \mTii requirements for each of the eight analysis bins.}
\centering
\renewcommand{\arraystretch}{1.2}
\begin{tabular}{c c c c}
Bin no. & \ptjt [\GeVns{}] & \ptmiss [\GeVns{}] & \mTii [\GeVns{}] \\
\hline
1 & 30--50 & 120--250 & ${<}100$ \\
2 & 30--50 & 120--250 & ${>}100$ \\
3 & 30--50 & ${>}250$ & \NA \\
4 & 50--100 & 120--250 & ${<}100$ \\
5 & 50--100 & 120--250 & ${>}100$ \\
6 & 50--100 & ${>}250$ & \NA \\
7 & ${>}100$ & \NA & ${<}100$ \\
8 & ${>}100$ & \NA & ${>}100$ \\
\hline
\end{tabular}
\label{tab:bins}
\end{table}

\section{Background estimation}
\label{sec:background}

The dominant background in this search arises from light quark and gluon jets that are misidentified as \distauh candidates.
Such jets can be produced in SM processes from $\Pp\Pp$ collisions, or in noncollision processes, such as beam halo interactions.
The latter contribution is greatly mitigated with the selections discussed in Section~\ref{sec:selection}.
Jets arising from beam halo are mostly trackless and preferably populate the forward (high $\abs{\eta}$) regions.
Any residual contamination from beam halo is expected to be negligible as \distau relies heavily on the properties of charged PF candidates in jets, and jets are required to have at least one charged PF candidate (\ie a track) and be reconstructed within $\abs{\eta} < 2.1$, as discussed in Section~\ref{sec:reco_obj}.
The contribution to the background yield from electrons, muons, and \tauh candidates from SM processes being misidentified as \distauh candidates is estimated from simulation to be negligible, \ie, with an event yield lower than 0.1 across all bins defined in Table~\ref{tab:bins}, because of the requirements on veto-electrons and veto-muons, and on the \dxy of the jets, as described in Sections~\ref{sec:reco_obj} and~\ref{sec:selection}.
Therefore, they are not included in the background estimation.

Simulated samples are not suitable for estimating the misidentified-jet background from SM processes owing to the limited number of events entering the SR, especially in the tails of the search variable distributions presented in Section~\ref{sec:selection}.
Therefore, we employ a technique fully based on control samples in data for evaluating the SM background arising from jets that are misidentified as \distauh candidates.
The procedure has been developed by combining the strategies used in Refs.~\cite{CMS:2022syk, CMS:2022wjc, CMS:2023yzg}.
In order to describe the method, the following quantities are to be considered:
\begin{itemize}
    \item Misidentification probability $f$, defined as the probability of a jet passing WPT, provided it has passed WPL:
    \begin{equation}
        f = n_\text{jet}(\text{WPT}) / n_\text{jet}(\text{WPL}) .
    \label{eq:fakerate}
    \end{equation}
    The probability of a jet passing WPL but failing WPT is then $1-f$.
    \item $N$: total number of events where exactly two jets pass WPL.
    \item $N_{\rtz}$, $N_{\rto}$, and $N_{\rtt}$: number of events satisfying \rtz, \rto, and \rtt criteria, respectively, as described in Table~\ref{tab:region_defs}.
\end{itemize}

Noting the above definitions, the following equations can be established:
\begin{subequations}
\begin{align}
N_{\rtz} &= (1-f_{1}) (1-f_{2}) N ,\\
N_{\rto} &= [f_{1} (1-f_{2}) + (1-f_{1}) f_{2}] N ,\\
N_{\rtt} &= f_{1} f_{2} N .
\end{align}
\label{eq:fake_bkg_1}%
\end{subequations}
Here $f_{1}$ and $f_{2}$ are the misidentification probabilities of the \distauh candidate with higher and lower \pt, respectively.
Taking the ratio between the subequations of Eq.~\eqref{eq:fake_bkg_1}, the following quantities can be obtained:
\begin{itemize}
    \item Prediction of $N_{\rto}$ from $N_{\rtz}$:
    \begin{equation}
        N_{\rtofz} \equiv N^\text{pred}_{\rto} = \frac{f_{1}(1-f_{2}) + (1-f_{1})f_{2}}{(1-f_{1})(1-f_{2})} N_{\rtz} .
        \label{eq:fake_rt1f0}
    \end{equation}
    
    \item Prediction of $N_{\rtt}$ from $N_{\rtz}$:
    \begin{equation}
        N_{\rttfz} \equiv N^\text{pred}_{\rtt} = \frac{f_{1} f_{2}}{(1-f_{1})(1-f_{2})} N_{\rtz}
        \label{eq:fake_rt2f0} .
    \end{equation}

    \item Prediction of $N_{\rtt}$ from $N_{\rto}$:
    \begin{equation}
        N_{\rttfo} \equiv N^\text{pred}_{\rtt} = \frac{f_{1} f_{2}}{f_{1} (1-f_{2}) + (1-f_{1}) f_{2}} N_{\rto} .
        \label{eq:fake_rt2f1}
    \end{equation}
\end{itemize}

Equation~\eqref{eq:fakerate} is used to measure the misidentification probability \frw from data, in the WCR described in Section~\ref{sec:selection}.
An alternate measurement, \frdy, is performed in the DYCR in data as a cross-check, and is found to agree with \frw within the statistical uncertainties.

The misidentified-jet background predictions in BRT1 and the SR are obtained using \frw para\-metrized in two dimensions, in bins of \pt (30--40, 40--50, 50--70, 70--120, and ${>}120$\GeV) and $\abs{\dxy}$ (0.2--0.5, 0.5--1, 1--2, 2--4, 4--10, 10--20, and ${>}20$\cm) of the jet.
Their values are, approximately, in the 10--20 and 5--10 (5--10 and 30--40)\% ranges with increasing \pt ($\abs{\dxy}$), depending on the data-taking period.

Jets with large \dxy can have a degraded \pt resolution as the displacement of the tracks from the primary vertex can affect the measurement of their transverse momenta.
Such jets would increase the \ptmiss in the event, potentially causing the event to fail the \mT requirement used to select the WCR.
This effect would induce a downward bias in the measured value of \frw, as the contribution from jets with large \dxy would be underestimated.
To test the impact of this possible bias, the measurement of \frw is also performed in the WCR, without imposing the aforementioned selection on \mT.
Additional measurements of \frw are performed in two subregions of this modified WCR, one with $50 \le \ptmiss < 120$\GeV, and the other with $\ptmiss \ge 120\GeV$.
These estimates of \frw obtained without the \mT requirement are found to be consistent with the nominal measurement reported above within the statistical uncertainties, even in the bins populated by jets with high \pt or large \dxy.
This indicates that the potential bias from this effect is negligible.

\begin{figure}[!ht]
    \centering
    \includegraphics[width=0.75\linewidth]{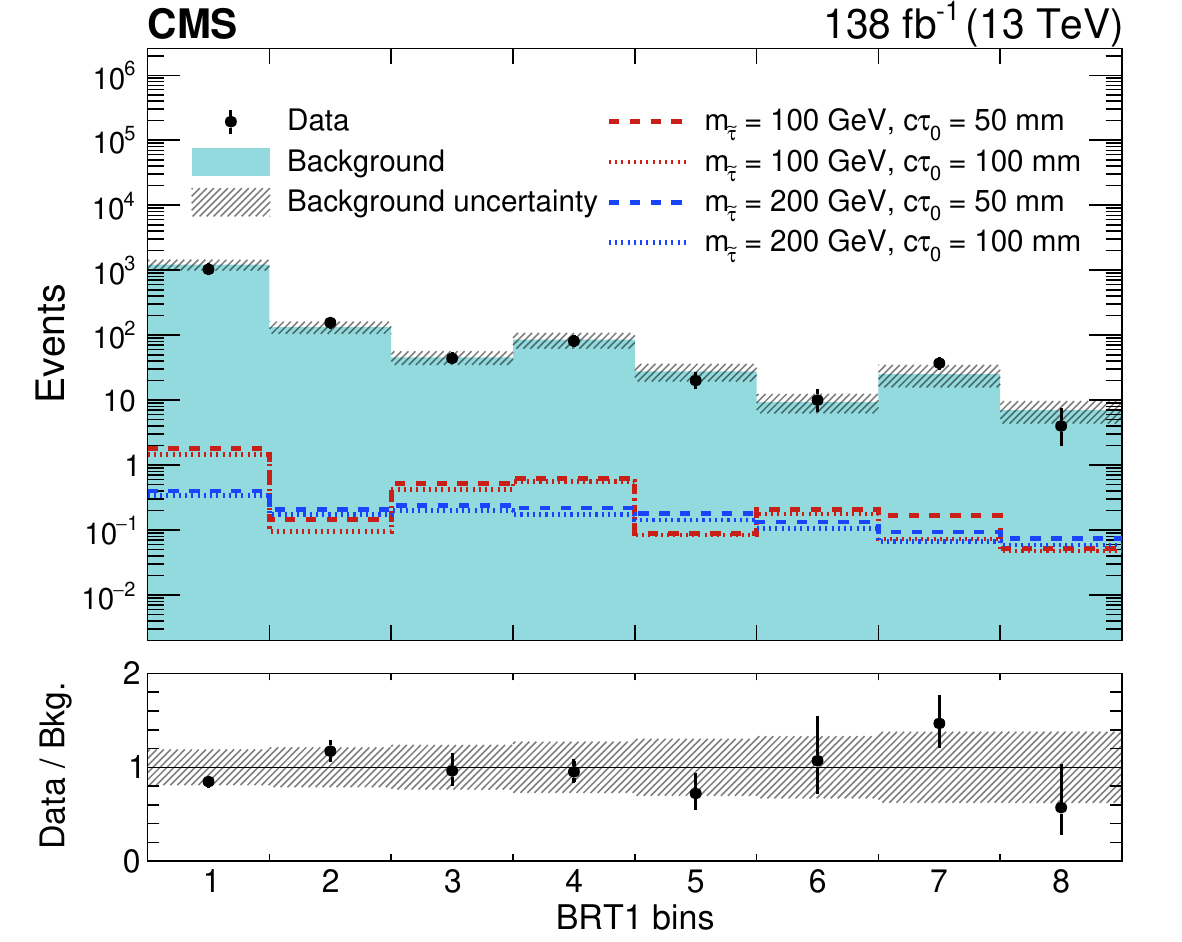}
    \caption{
    Observed and predicted event yields in the eight BRT1 bins, as defined in Table~\ref{tab:bins}.
    The signal yields expected in the maximally mixed scenario for four representative sets of $(\mstau~[{\GeV}],\ \ctauz~[{\mm}])$ values are overlaid: (100, 50), (100, 100), (200, 50), and (200, 100).
    The lower panel indicates the ratio of the observed number of events to the total predicted number of background events in each bin.
    The shaded bands indicate the statistical and systematic uncertainties in the background, added in quadrature.
    }
    \label{fig:yields_brt1}
\end{figure}

We use BRT1 as the primary region for validating the background estimation as it is adjacent to the SR, while having a negligible signal contamination of ${\lesssim}1\%$, compared to the predicted background.
Figure~\ref{fig:yields_brt1} shows the observed event yields in BRT1 compared with the signal and background predictions, the latter (obtained using $N_{\rtofz}$) being found to be consistent with the observed data.
Additional validations are performed by using \frw (\frdy) to predict the yields in DYCRT1 and DYCRT2 (WCRT1 and WCRT2) and comparing them with the observed data in the corresponding regions.
These comparisons are also found to agree within the uncertainties.

The primary background prediction in the SR is obtained using $N_{\rttfo}$ (Eq.~\ref{eq:fake_rt2f1}).
An alternate measurement is performed using $N_{\rttfz}$ (Eq.~\ref{eq:fake_rt2f0}), serving as a cross-check against potential signal contamination in BRT1, which would have an impact on the estimation of $N_{\rttfo}$.
The two estimates are found to agree within uncertainties, and the difference between them is propagated as a systematic uncertainty, as reported in Section~\ref{sec:systematics}.

\section{Systematic uncertainties}
\label{sec:systematics}

Several sources of systematic uncertainties affecting the prediction of both signal and background yields are taken into account, and their contributions are summarized in Table~\ref{tab:systs}.

\begin{table}[!ht]
\topcaption{
Relative systematic uncertainties expressed as a percentage (\%) of the signal and background yields in the SR, from various sources considered in this search, after accounting for their correlations between the data-taking periods.
The three values correspond to the minimum, median, and maximum values across the eight SR bins, as defined in Table~\ref{tab:bins}.
For cases where the minimum and maximum values differ by less than 1, only the median is shown.
In the header row, \mstau and \ctauz are in units of \GeV and \mm, respectively.
The uncertainty values shown here are prior to the maximum likelihood fit described in Section~\ref{sec:results}.
}
\renewcommand{\arraystretch}{1.4}
\cmsTable{\begin{tabular}{lccccc}

Uncertainty source & Background & \centerstack{$m_{\PSGt}(100)$\\$\ctauz(50)$} & \centerstack{$m_{\PSGt}(100)$\\$\ctauz(100)$} & \centerstack{$m_{\PSGt}(200)$\\$\ctauz(50)$} & \centerstack{$m_{\PSGt}(200)$\\$\ctauz(100)$} \\[\cmsTabSkip]
\hline
\centerstack[l]{Parton distribution\\[-0.25\cmsTabSkip] function}\rule{0pt}{1.1\normalbaselineskip}  & \NA & $3.5,\ 3.8,\ 6$ & $2.9,\ 4.2,\ 6.4$ & $4,\ 4.6,\ 5.9$ & $3.9,\ 4.4,\ 5.2$ \\[0.5\cmsTabSkip]
$\mu_{\text{R}}$ scale & \NA & ${<}0.1,\ 0.29,\ 1.4$ & ${<}0.1,\ 0.46,\ 1.3$ & ${<}0.1$ & $0.1$ \\
$\mu_{\text{F}}$ scale & \NA & $0.62,\ 4.3,\ 7.1$ & $1.3,\ 5.1,\ 7.1$ & $0.63,\ 1.1,\ 2.9$ & $0.65,\ 0.94,\ 2.8$ \\
Jet energy scale & \NA & $0.83,\ 2.4,\ 7.4$ & $0.68,\ 3.2,\ 10$ & $0.39,\ 1.4,\ 2.5$ & $0.47,\ 1.1,\ 2.5$ \\
Jet energy resolution & \NA & $0.19,\ 0.54,\ 7.4$ & $0.14,\ 0.82,\ 4.3$ & $0.22$ & $0.17$ \\[0.5\cmsTabSkip]
\centerstack[l]{$\ptmiss$ unclustered\\[-0.25\cmsTabSkip] energy} & \NA & $0.26,\ 0.9,\ 7.4$ & $0.31,\ 1.1,\ 8.9$ & $0.28$ & $0.33$ \\[0.5\cmsTabSkip]
Pileup & \NA & $0.66,\ 1.2,\ 3.3$ & $0.62,\ 0.98,\ 5.1$ & $0.61$ & $0.59$ \\
Integrated luminosity & \NA & $1.7$ & $1.7$ & $1.7$ & $1.7$ \\
$\ptmiss$ trigger & \NA & $1.6,\ 16,\ 20$ & $1.6,\ 15,\ 20$ & $1.7,\ 14,\ 16$ & $1.6,\ 14,\ 16$ \\[0.5\cmsTabSkip]
\distau identification & \NA & $17,\ 18,\ 19$ & $17,\ 18,\ 19$ & $17$ & $18$ \\[\cmsTabSkip]
\centerstack[l]{Misid. probability\\[-0.25\cmsTabSkip] estimation region} & $11,\ 11,\ 13$ & \NA & \NA & \NA & \NA \\[1.5\cmsTabSkip]
\centerstack[l]{Misid. probability\\[-0.25\cmsTabSkip] sample size} & $15,\ 30,\ 57$ & \NA & \NA & \NA & \NA \\[1.5\cmsTabSkip]
\centerstack[l]{Alternative bkg.\\[-0.25\cmsTabSkip] prediction} & $8.5,\ 24,\ 53$ & \NA & \NA & \NA & \NA \\[1.5\cmsTabSkip]
\centerstack[l]{Bkg. correction \\[-0.25\cmsTabSkip] from BRT1} & $2.8,\ 5.2,\ 8.1$ & \NA & \NA & \NA & \NA \\[0.5\cmsTabSkip]
\hline
\end{tabular}}

\label{tab:systs}
\end{table}

Sources of theoretical uncertainties affecting the signal simulation include the PDFs and missing higher-order corrections.
The effect of these are estimated from the variation in signal yields from alternate PDF sets~\cite{NNPDF:2014otw, Butterworth:2015oua, NNPDF:2017mvq}, and from independently varying the factorization ($ \mu_{\text{F}} $) and renormalization ($ \mu_{\text{R}} $) scales up and down by a factor of two with the constraint $0.5\leq(\mu_{\text{R}}/\mu_{\text{F}})\leq2$ to account for the uncertainty arising from missing higher-order corrections.
The \textsc{SysCalc} package~\cite{Kalogeropoulos:2018cke} is used for this purpose.

The experimental sources affecting the signal include the jet energy scale and jet energy resolution, the effect of unclustered energy in calculating \ptmiss, pileup reweighting, integrated luminosity, and the uncertainty in the modeling of the \ptmiss trigger.
The last uncertainty is taken to be the difference from unity of the simulation-to-data correction to the trigger efficiency for events with $\ptmiss<250\GeV$, estimated with events selected by a single-muon trigger.
This uncertainty is larger than the statistical uncertainty in the correction factor in this \ptmiss region, and is intended to account for potential mismodeling of the trigger efficiency turn-on curve in simulation.
For events with higher \ptmiss, \ie in the plateau region of the trigger efficiency, the statistical uncertainty in the correction factor is used.
The simulation is reweighted to make its pileup vertex distribution identical to that of the data.
The uncertainty in estimating the number of pileup interactions is estimated by varying the total inelastic cross section by $\pm$4.6\%~\cite{Sirunyan:2018nqx}.
This is propagated as an uncertainty in the pileup reweighting factor applied to the simulation.
The uncertainty in the measured integrated luminosity amounts to 1.2, 2.3, and 2.5\% in 2016, 2017, and 2018~\cite{CMS-LUM-17-003, CMS-PAS-LUM-17-004, CMS-PAS-LUM-18-002}, respectively.
In addition, the uncertainty in the correction factor for the \distau identification efficiency described in Section~\ref{sec:reco_distau}, is also taken into account.

All sources of systematic uncertainties affecting the signal are considered uncorrelated across data-taking years, except for those from theory and the integrated luminosity uncertainty.
The theory uncertainties are treated as fully correlated across the years, and the integrated luminosity uncertainty is treated as partially correlated, which results in a value of 1.7\% on combining the years.

The following systematic uncertainties are propagated to the prediction of the final background yields.
They are discussed here in the order of their appearance in Table~\ref{tab:systs}.

Since the predicted background in the BR is obtained using $\frw$, which is measured in a $\wjets$ enriched region, one needs to account for how differences between WCR and BR can impact the extrapolation of the misidentification probability in the BR ($\frbr$).
These can arise from differences between the two regions in the flavor composition of the parton corresponding to the jet that is misidentified as a $\distauh$ candidate.
We use simulations of the SM processes described in Section~\ref{sec:simulation} to estimate the relative difference $\abs{\frw^{\mc}-f^{\mc}_\text{BR}}/\frw^{\mc}$, which is found to be about 10\%.
This is treated as an uncertainty in $\frw$ (labeled ``Misid. probability estimation region'') correlated across data-taking years and propagated to the background prediction.
The statistical uncertainty in $\frw$ due to the finite size of the WCR (labeled ``Misid. probability sample size'') is also propagated to the prediction of the background yields.
In addition, the difference between the primary and alternate background predictions, as discussed in Section~\ref{sec:background} is taken as an uncertainty (labeled ``Alternative bkg. prediction'').
These last two sources of uncertainties are treated as uncorrelated across data-taking years.

The observed data in the BRT1 region for the first $19\fbinv$ of 2016 data taking were observed to be lower than the estimated background by a factor of 0.7 uniformly across all bins.
The discrepancy can possibly be attributed to the saturation effects observed in the APV25 readout chip preamplifier used in the tracker~\cite{CMS:2025bbk}, during the same data-taking period.
Therefore, an additional correction of $0.7 \pm 0.3$ is applied to the SR in that portion of the data set, with the uncertainty (``Bkg. correction from BRT1'') being taken as the deviation of the factor from unity.
This correction has a negligible effect on the final results described in Section~\ref{sec:results}.

\section{Results}
\label{sec:results}

\begin{figure}[!htbp]
    \centering
    \includegraphics[width=0.7\linewidth]{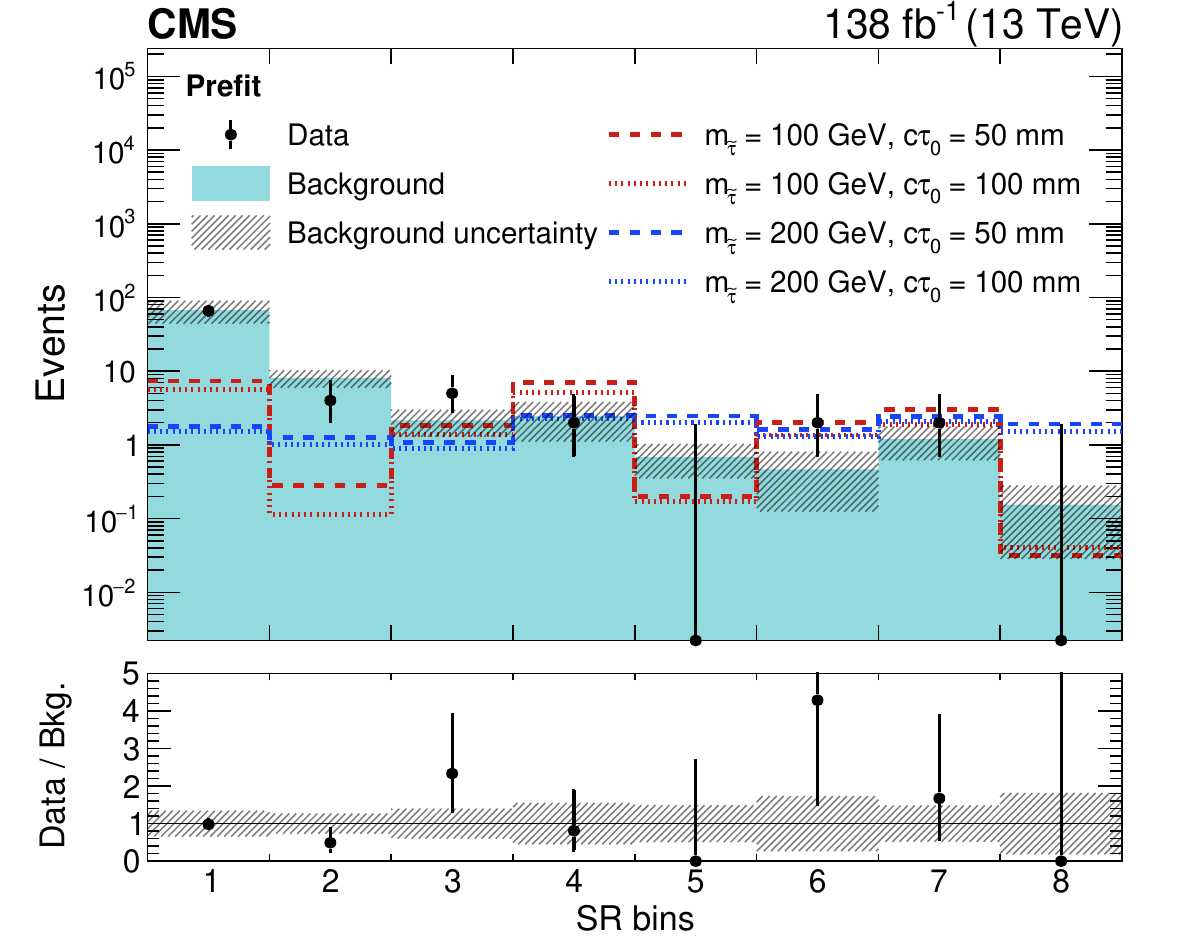}
    \includegraphics[width=0.7\linewidth]{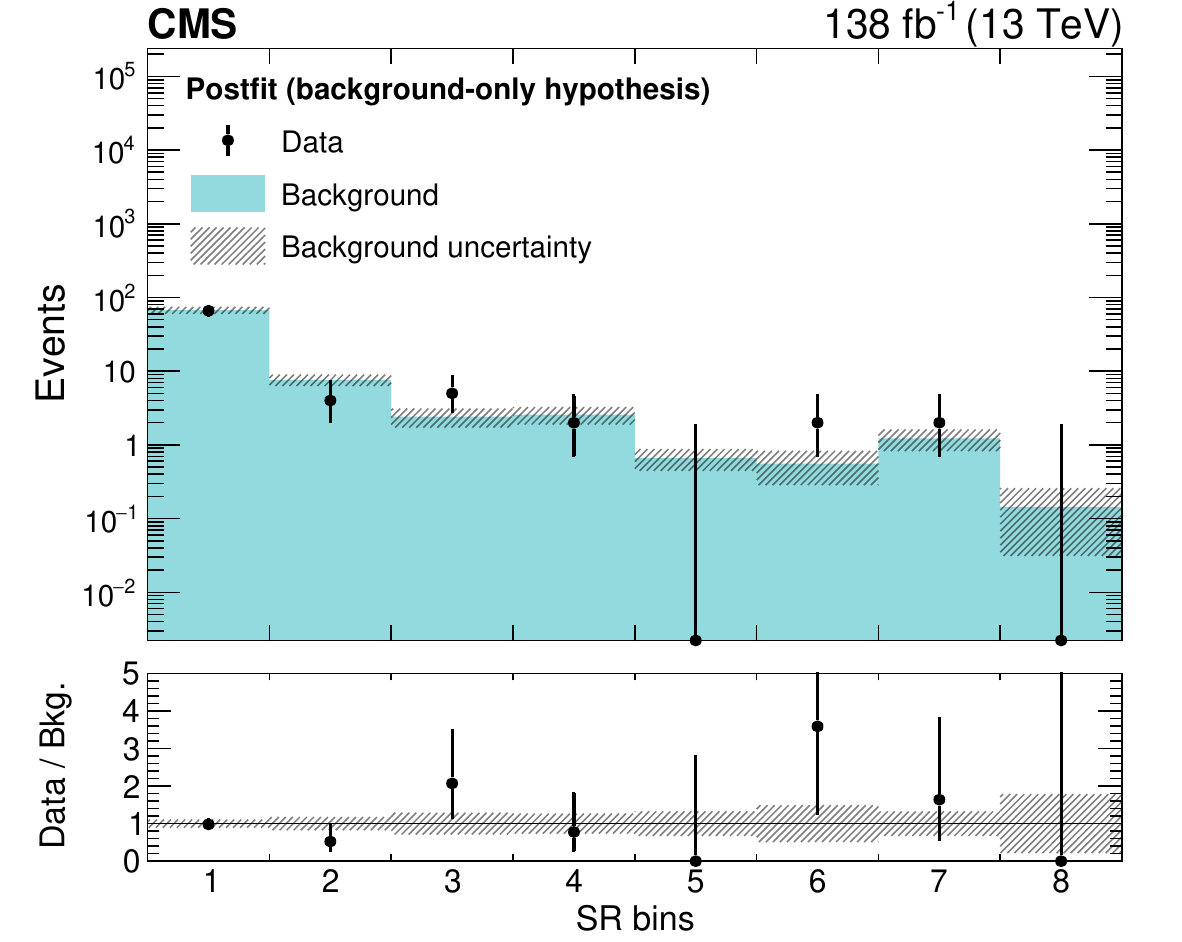}
    \caption{
    Observed and predicted event yields in the eight SR bins as defined in Table~\ref{tab:bins}.
    The signal yields expected in the maximally mixed scenario for four representative sets of $(\mstau~[{\GeVns}],\ \ctauz~[{\mmns}])$ values are overlaid: (100, 50), (100, 100), (200, 50), and (200, 100).
    In bins where the observed yield is zero, the Garwood interval at 68\% \CL is shown as a positive uncertainty.
    The lower panel indicates the ratio of the observed number of events to the total predicted number of background events in each bin.
    The shaded bands indicate the statistical and systematic uncertainties in the background prediction, added in quadrature.
    The predicted yields and uncertainties shown in the upper (lower) panel are before (after) the maximum likelihood fit to data under the background-only hypothesis, as described in Section~\ref{sec:results}.
    }
    \label{fig:yields_sr}
\end{figure}

The observed and expected yields along with their uncertainties in all eight SR bins are shown in Table~\ref{tab:yields_sr} and Fig.~\ref{fig:yields_sr}, along with four representative signal hypotheses.
The number of events observed in data is found to be consistent with the SM background prediction.

\begin{table}[!htb]
\topcaption{
Predicted background and signal yields in the eight SR bins, as defined in Table~\ref{tab:bins}, are shown alongside the background yields obtained from the maximum likelihood fit to data under the background-only hypothesis and the number of events observed in the recorded data.
All yields, except for the number of recorded events, are quoted with uncertainties obtained by adding statistical and systematic components in quadrature. Signal and predicted background uncertainties correspond to the prefit values, while the postfit background uncertainties are taken from the fit described in Section~\ref{sec:results}.
In the header row, \mstau and \ctauz are in units of \GeVns and \mmns, respectively.
}
\renewcommand{\arraystretch}{1.5}
\centering
\begin{tabular}{L{0.7cm}cC{1.7cm}C{1.7cm}C{2.1cm}C{2.1cm}C{1.9cm}C{1.9cm}}
SR bin & Data & Bkg. (prefit) & Bkg. (postfit) & $m_{\PSGt}(100)$ $\ctauz(50)$ & $m_{\PSGt}(100)$ $\ctauz(100)$ & $m_{\PSGt}(200)$ $\ctauz(50)$ & $m_{\PSGt}(200)$ $\ctauz(100)$ \\
\hline
1 & 66 & $67{\pm}23$ & $67{\pm}8$ & $7.3{\pm}1.8$ & $5.7{\pm}1.4$ & $1.8{\pm}0.4$ & $1.5{\pm}0.4$ \\
2 & 4 & $8.1{\pm}2.2$ & $7.6{\pm}1.4$ & $0.28{\pm}0.09$ & $0.11{\pm}0.04$ & $1.26{\pm}0.29$ & $1.02{\pm}0.23$ \\
3 & 5 & $2.1{\pm}0.9$ & $2.4{\pm}0.7$ & $1.8{\pm}0.4$ & $1.39{\pm}0.30$ & $1.08{\pm}0.21$ & $0.89{\pm}0.17$ \\
4 & 2 & $2.5{\pm}1.3$ & $2.6{\pm}0.7$ & $7.1{\pm}1.8$ & $5.2{\pm}1.3$ & $2.5{\pm}0.6$ & $2.2{\pm}0.5$ \\
5 & 0 & $0.7{\pm}0.4$ & $0.66{\pm}0.22$ & $0.20{\pm}0.06$ & $0.17{\pm}0.06$ & $2.5{\pm}0.6$ & $2.0{\pm}0.5$ \\
6 & 2 & $0.47{\pm}0.34$ & $0.56{\pm}0.25$ & $2.0{\pm}0.4$ & $1.36{\pm}0.30$ & $1.62{\pm}0.31$ & $1.29{\pm}0.25$ \\
7 & 2 & $1.2{\pm}0.6$ & $1.2{\pm}0.4$ & $3.0{\pm}0.7$ & $1.9{\pm}0.5$ & $2.4{\pm}0.5$ & $2.1{\pm}0.5$ \\
8 & 0 & $0.15{\pm}0.13$ & $0.14{\pm}0.10$ & $0.032{\pm}0.015$ & $0.041{\pm}0.017$ & $1.9{\pm}0.4$ & $1.5{\pm}0.4$ \\[0.5\cmsTabSkip]
Tot. & 81 & $83{\pm}27$ & $83{\pm}11$ & $22{\pm}5$ & $16{\pm}4$ & $15.1{\pm}3.3$ & $12.6{\pm}2.8$ \\
\hline
\end{tabular}
\label{tab:yields_sr}
\end{table}

The test statistic used for the interpretation of the result is the LHC-style profile likelihood ratio $q_{r} = -2 \ln{(\mathcal{L}_{r} / \mathcal{L}_{\text{max}})}$, where $ \mathcal{L}_{r} $ is the maximum likelihood for a fixed signal strength modifier $r$, and $ \mathcal{L}_{\text{max}} $ is the global maximum of the likelihood~\cite{CMS-NOTE-2011-005}.
The systematic uncertainties discussed in Section~\ref{sec:systematics} are modeled by log-normal distributions~\cite{CMS-NOTE-2011-005} in the likelihood function.
We set upper limits on signal production cross section at 95\% \CL using a modified frequentist approach with a \CLs criterion~\cite{JUNK1999435, Read_2002}, with test-statistic distributions obtained from ensembles of pseudo-experiments.
In this calculation, all the background and signal uncertainties are incorporated as nuisance parameters and profiled in the maximum likelihood fit~\cite{CMS-NOTE-2011-005}.

\begin{figure}[!htp]
    \centering
    \includegraphics[width=0.75\linewidth]{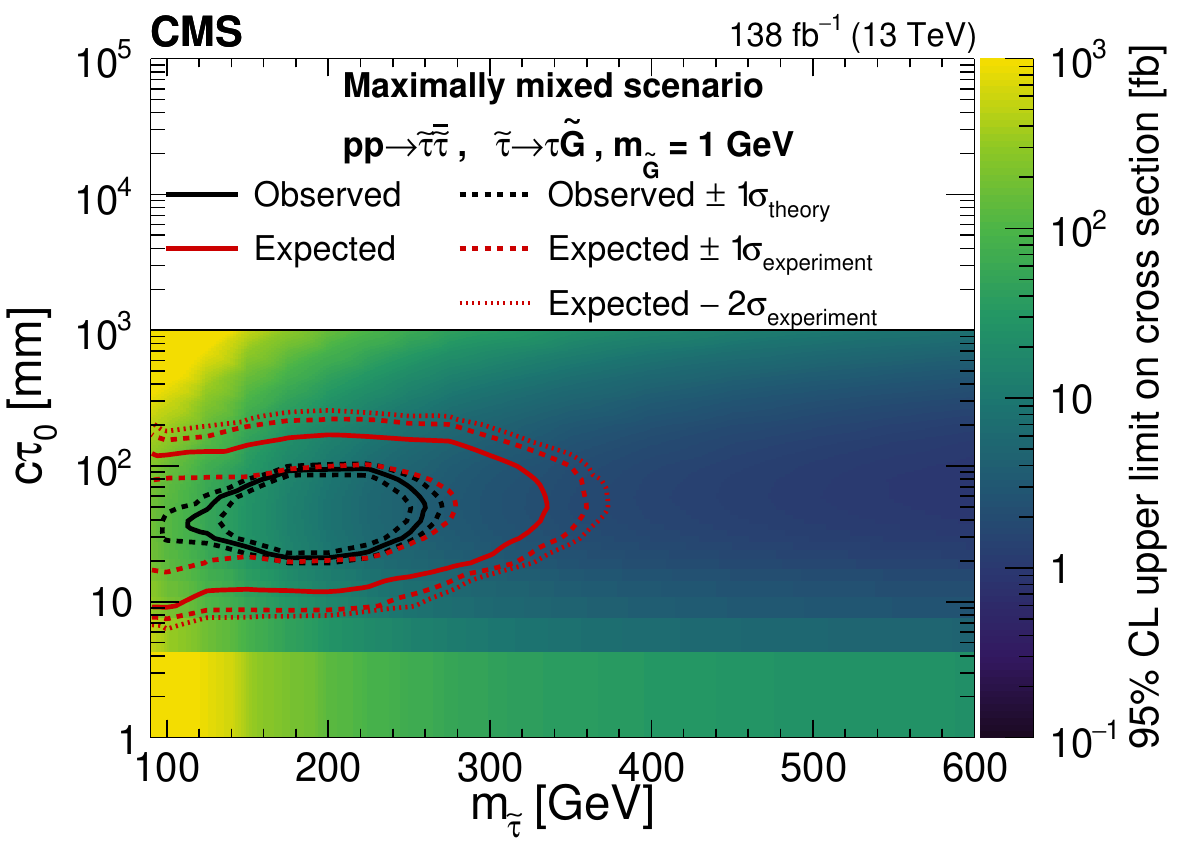}
    \includegraphics[width=0.75\linewidth]{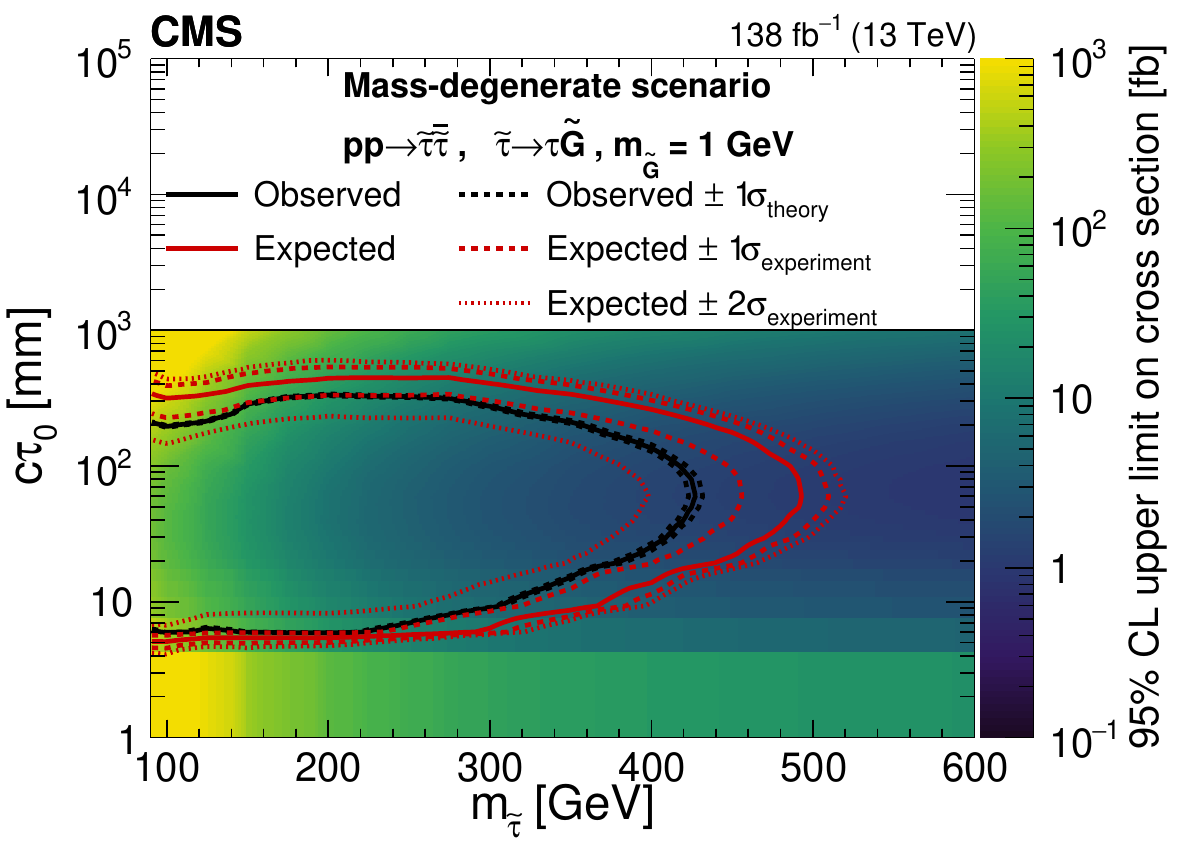}
    \caption{
    Exclusion limits at 95\% \CL for the pair production of long-lived staus decaying to a \PGt lepton and a nearly massless gravitino ($m_\PXXSG=1$\GeV), displayed in the plane of $\mstau$ and $\ctauz$.
    The maximally mixed (mass-degenerate) scenario is presented in the upper (lower) panel.
    The color axis represents the observed upper limit on the cross section.
    The signal cross sections are evaluated using NLO$+$NLL calculations.
    The black (red) lines represent the observed (expected) limits on $\mstau$ and $\ctauz$ and enclose the region excluded at 95\% \CL.
    The solid lines represent the central values.
    The dashed (dotted) red lines indicate the region containing 68\% (95\%) of the distribution of limits expected under the background-only hypothesis accounting for experimental uncertainties ($\sigma_\text{experiment}$).
    The red dotted line corresponding to ``Expected$+2\sigma_\text{experiment}$" is not shown for the maximally mixed scenario, as it does not exclude any points in this plane.
    The dashed black lines show the change in the observed limit from the variation of the signal cross sections within their theoretical uncertainties ($\sigma_\text{theory}$).
    }
    \label{fig:limits_2d}
\end{figure}

\begin{figure}[!htbp]
    \centering 
    \includegraphics[width=0.42\linewidth]{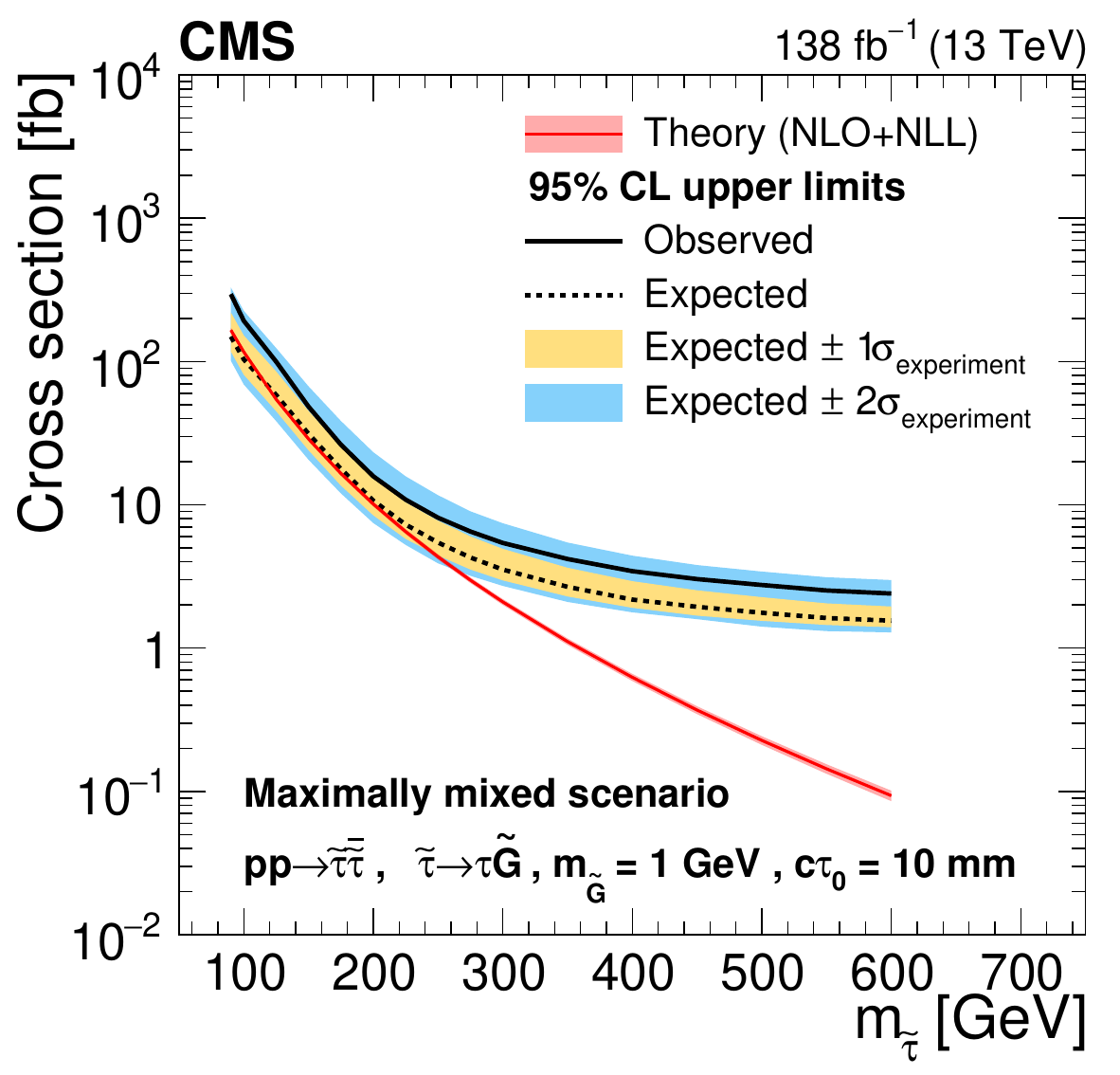}
    \includegraphics[width=0.42\linewidth]{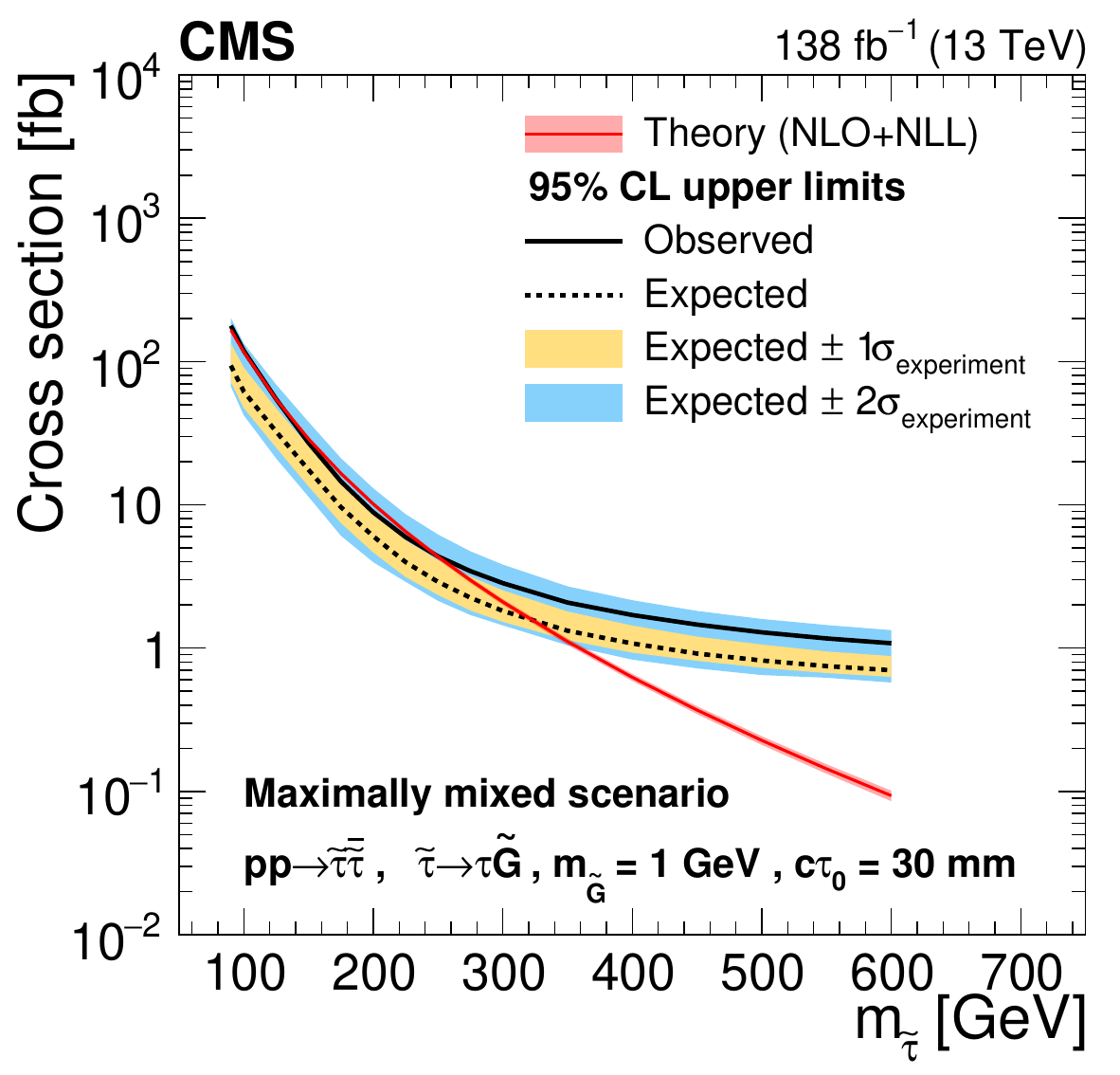}
    \\
    \includegraphics[width=0.42\linewidth]{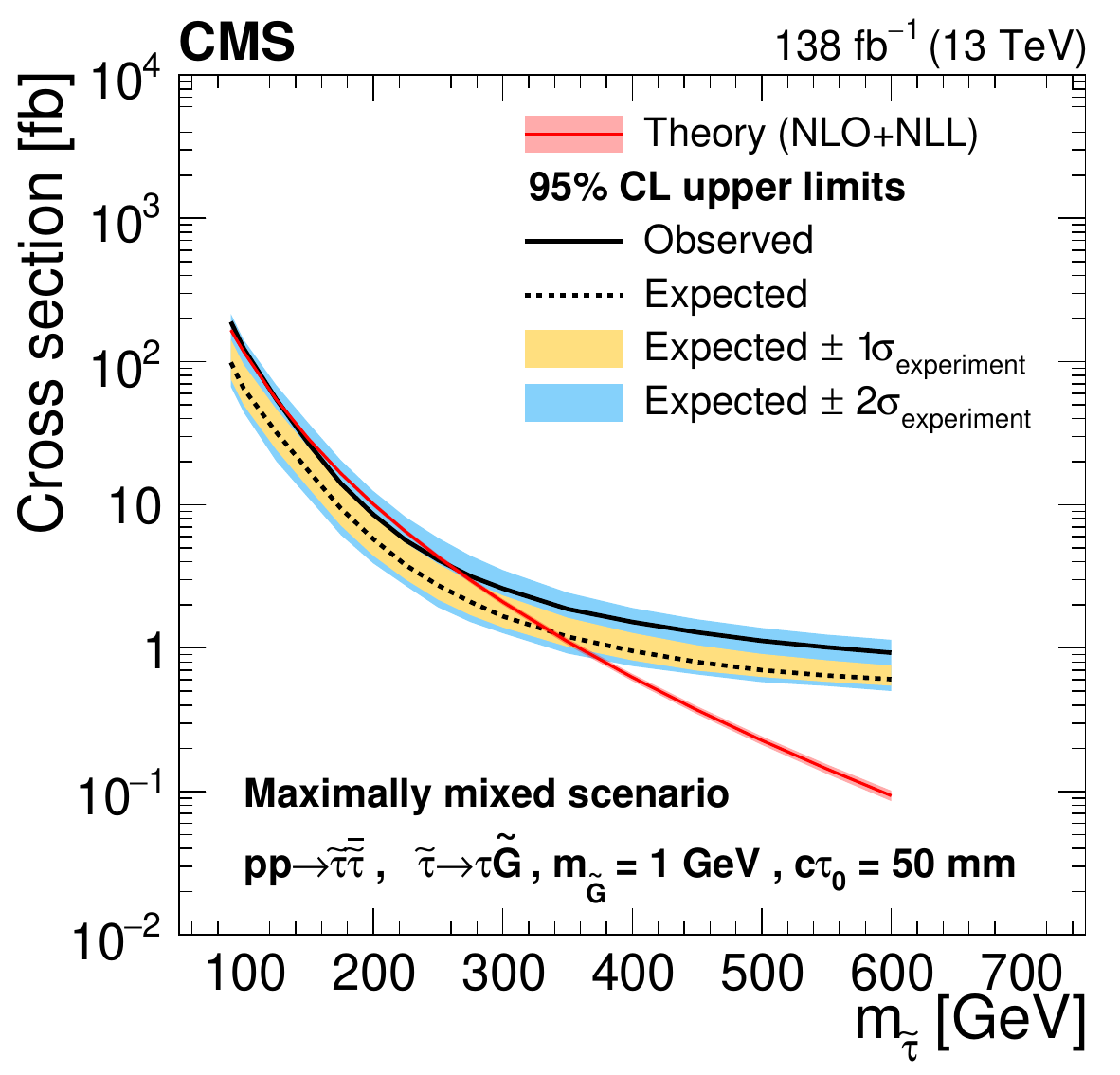}
    \includegraphics[width=0.42\linewidth]{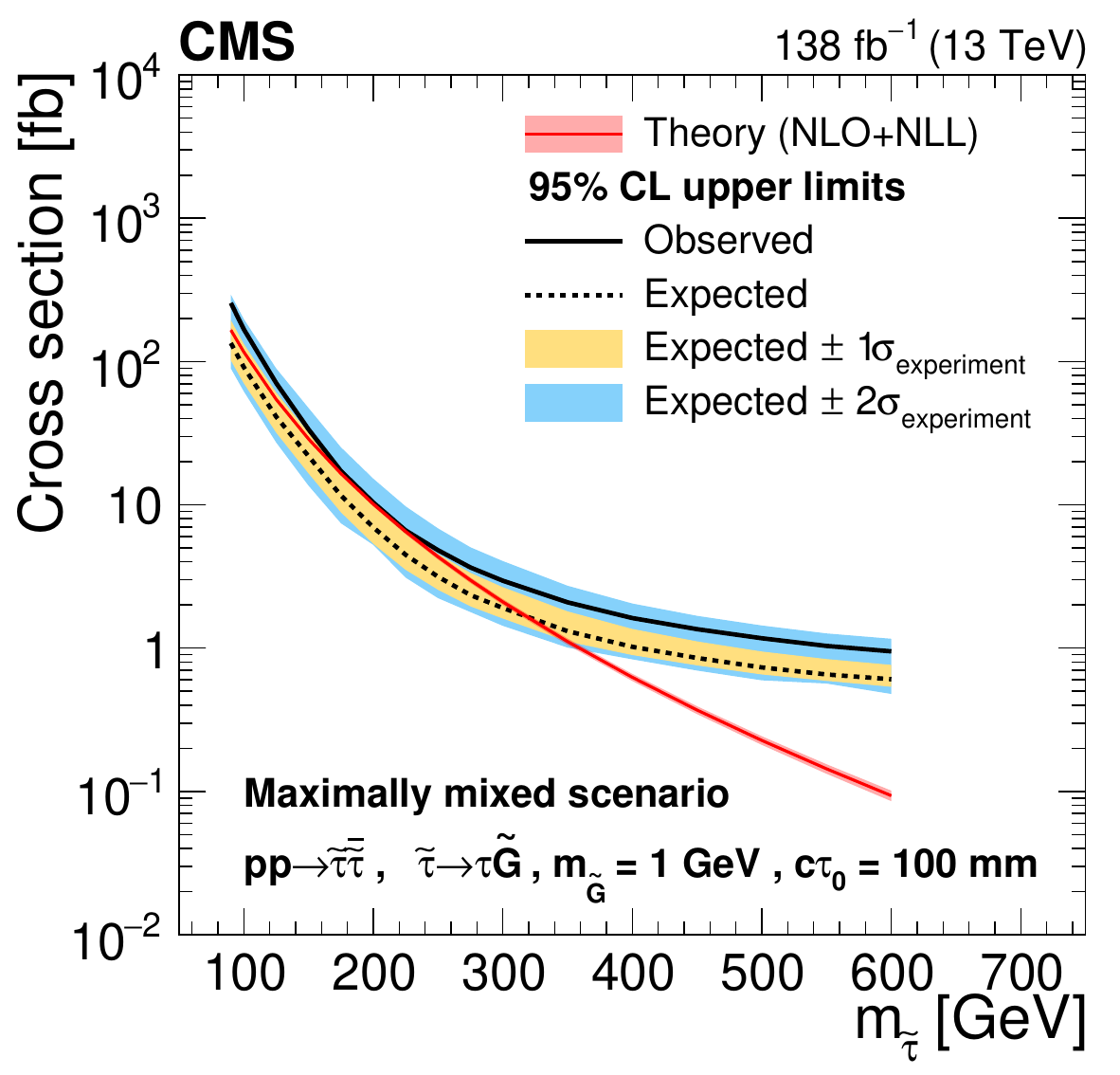}
    \\
    \includegraphics[width=0.42\linewidth]{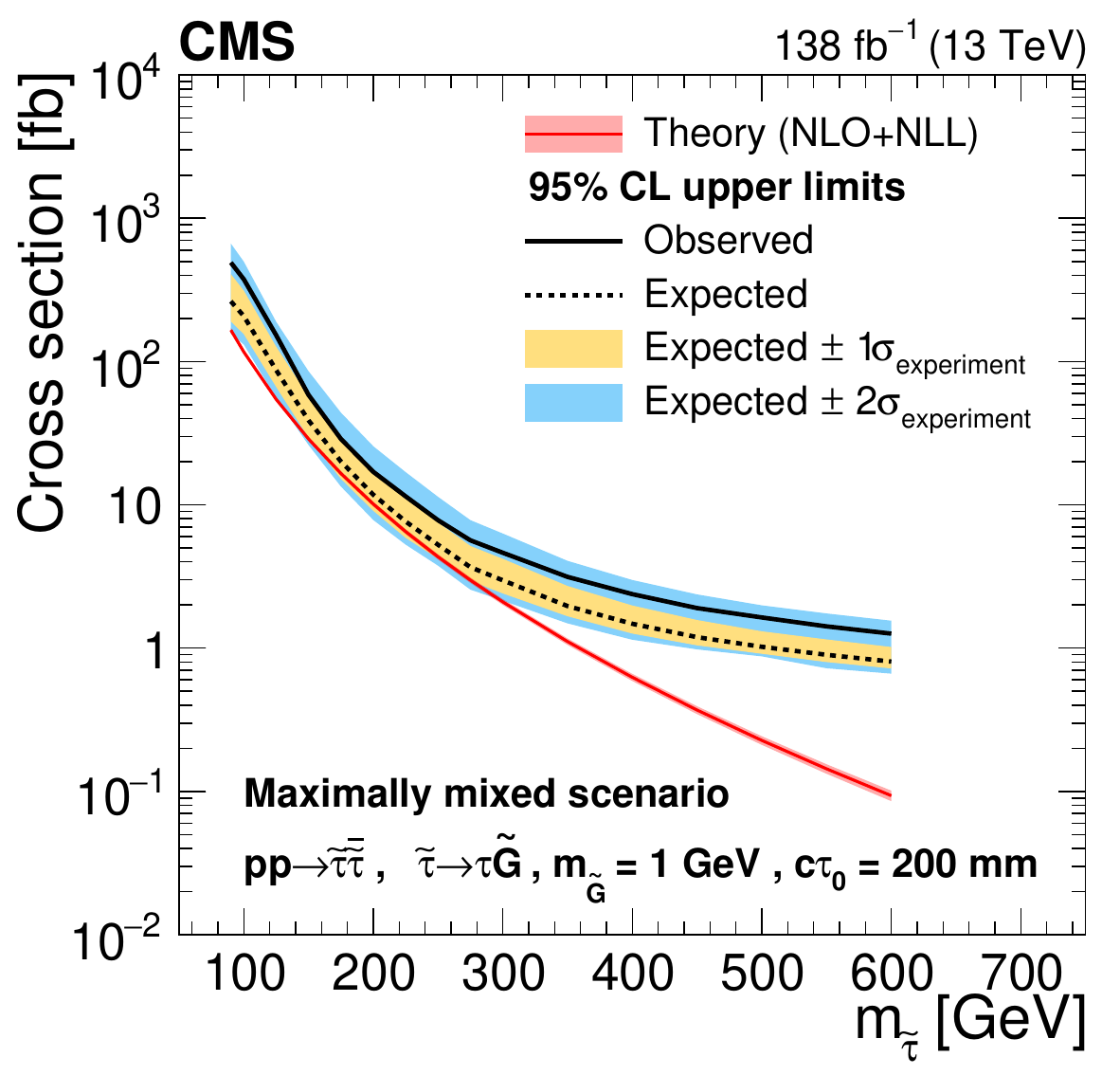}
    \includegraphics[width=0.42\linewidth]{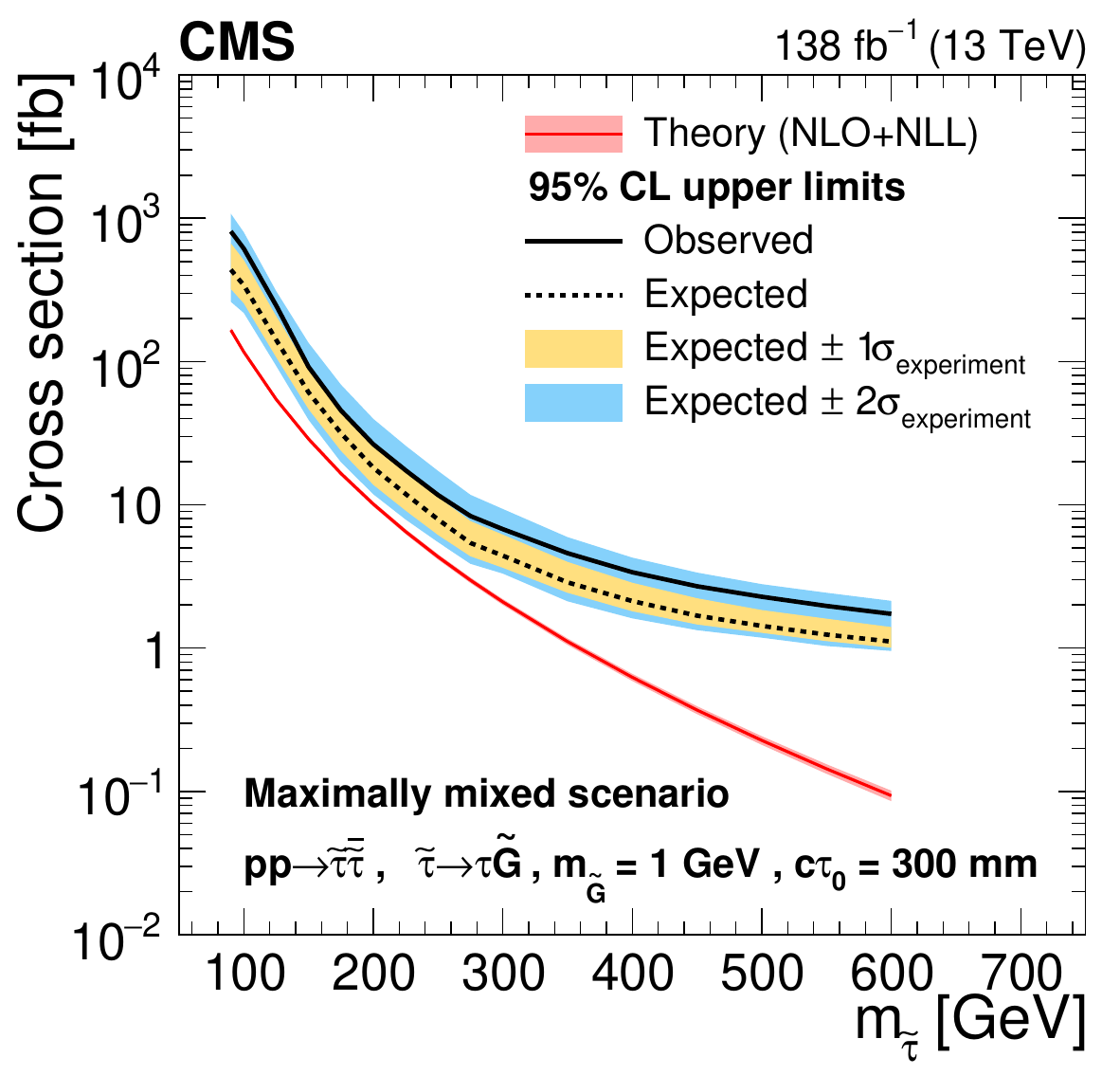}
    \caption{
    Cross section upper limits at 95\% \CL for the pair production of long-lived staus decaying to a \PGt lepton and a nearly massless gravitino ($m_\PXXSG=1$\GeV) in the maximally mixed scenario, as a function of $\mstau$, for $\ctauz=$ 10, 30, 50, 100, 200, and 300\mm (left to right, upper to lower).
    The inner (yellow) and outer (blue) bands indicate the regions containing 68 and 95\%, respectively, of the distribution of limits expected under the background-only hypothesis.
    The central values of the expected upper limits are denoted by the dashed black line.
    The solid black line represents the observed upper limits.
    The signal cross sections and uncertainties evaluated using NLO$+$NLL calculations are shown as a red line and shaded band.
    }
    \label{fig:limits_1d_mm}
\end{figure}

\begin{figure}[!htbp]
    \centering
    \includegraphics[width=0.42\linewidth]{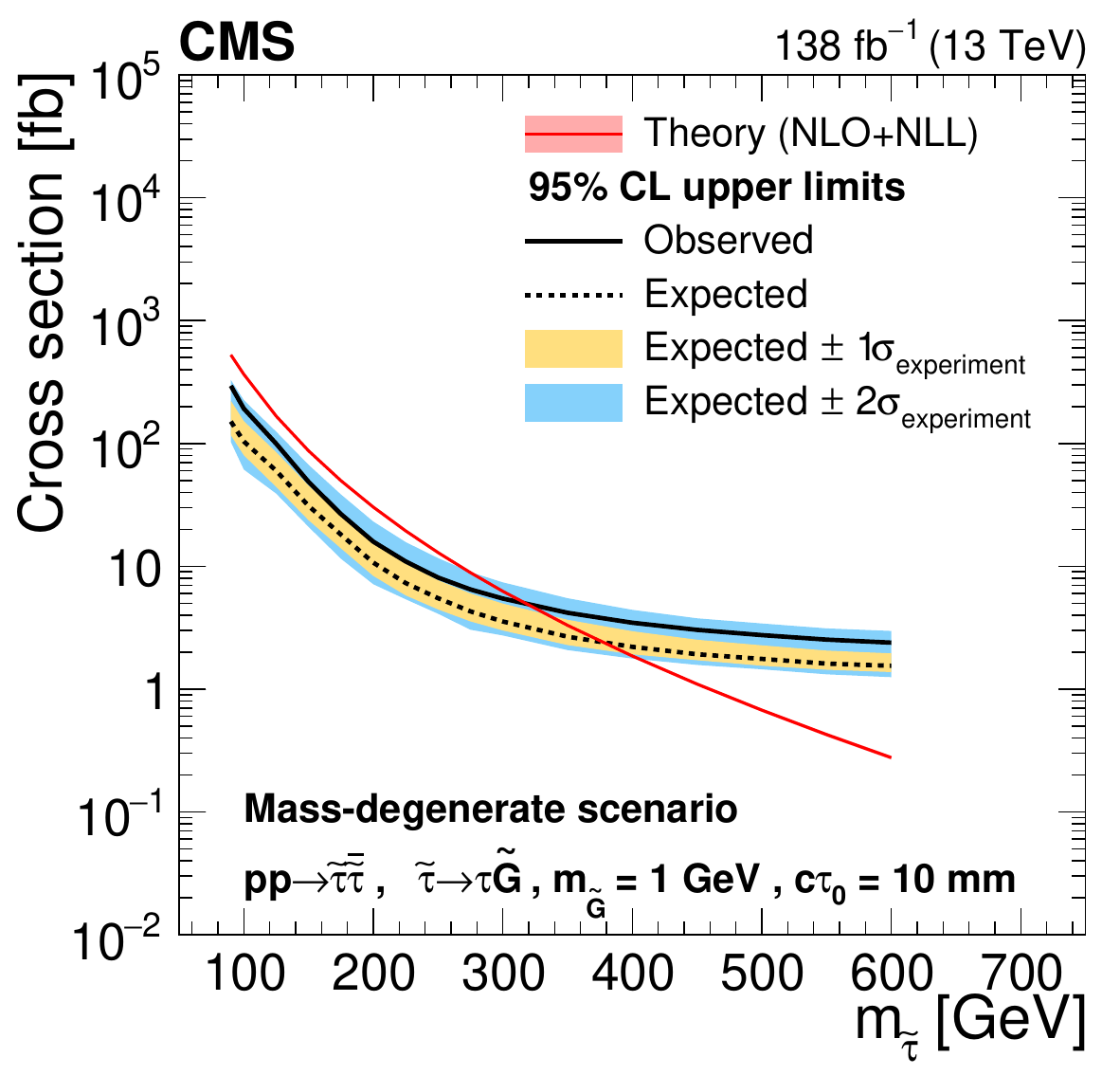}
    \includegraphics[width=0.42\linewidth]{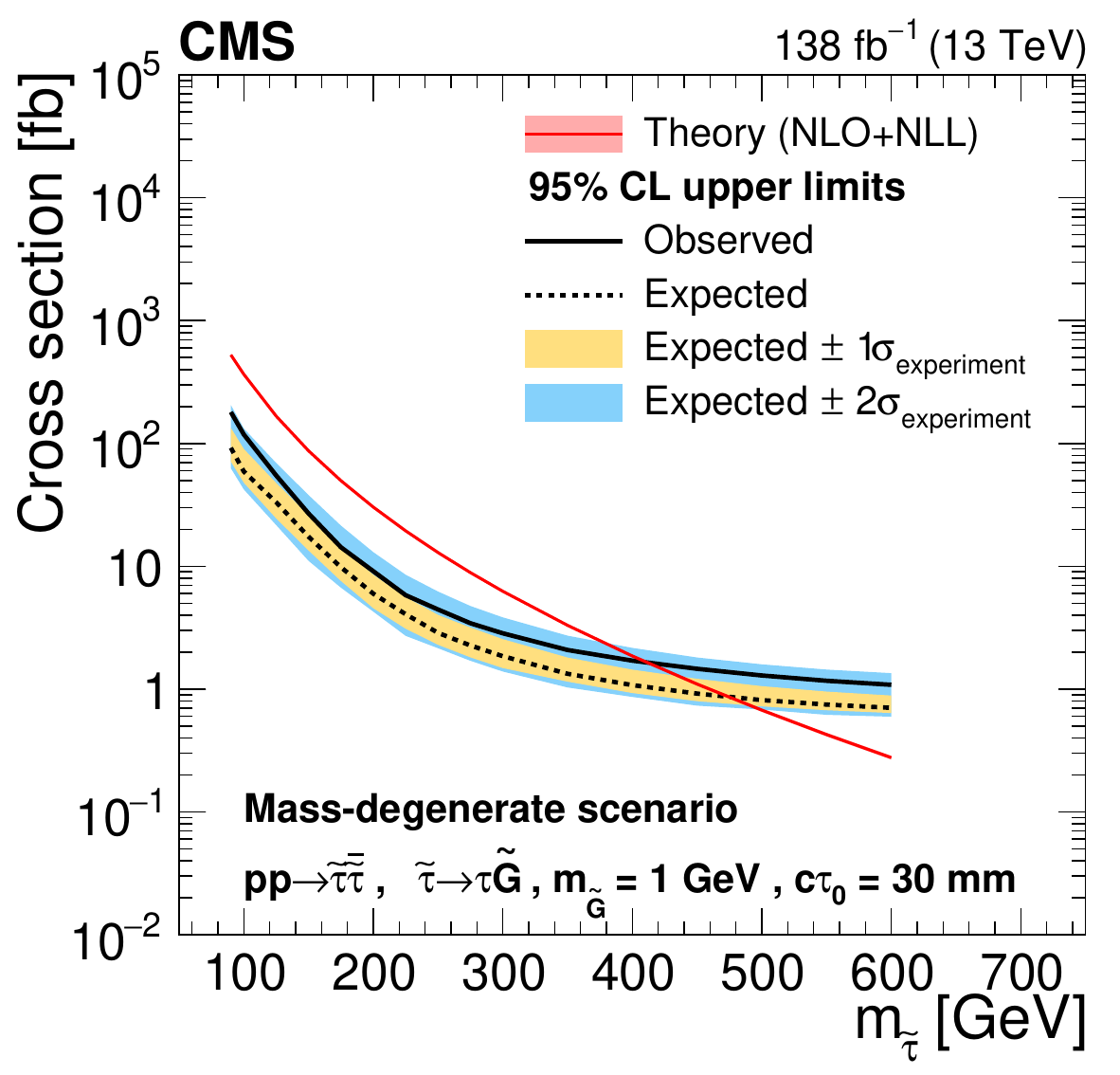}
    \\
    \includegraphics[width=0.42\linewidth]{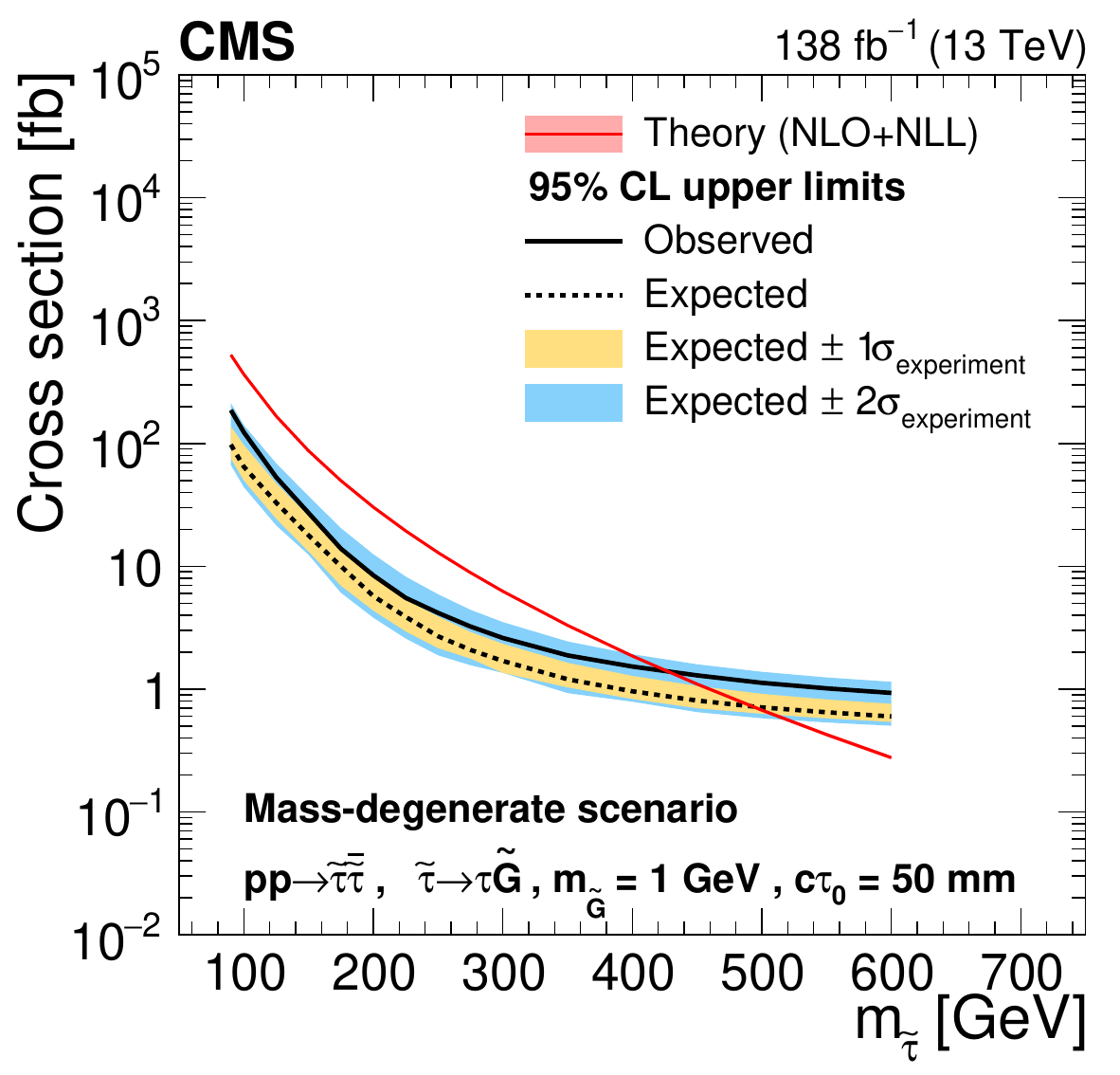}
    \includegraphics[width=0.42\linewidth]{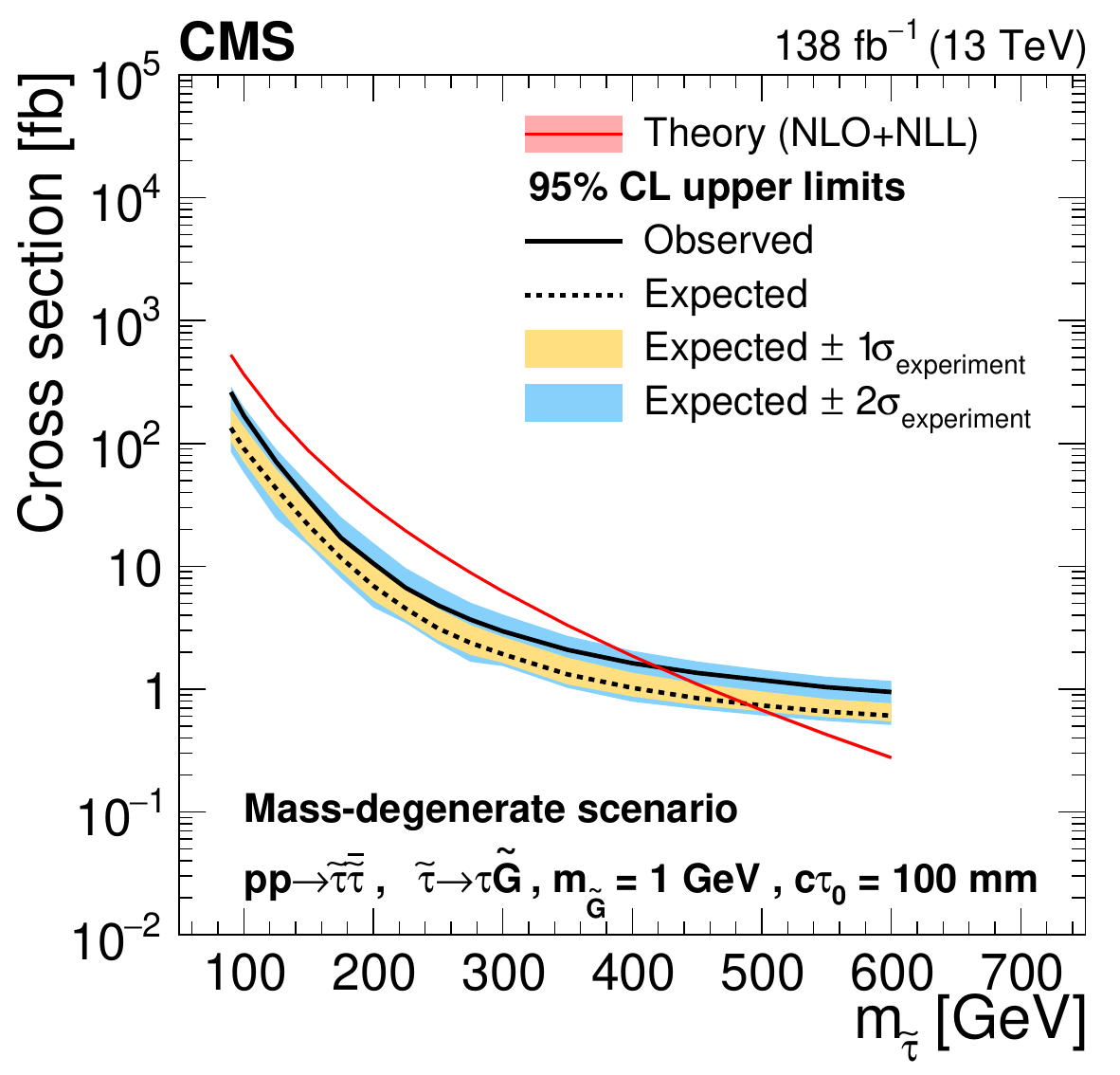}
    \\
    \includegraphics[width=0.42\linewidth]{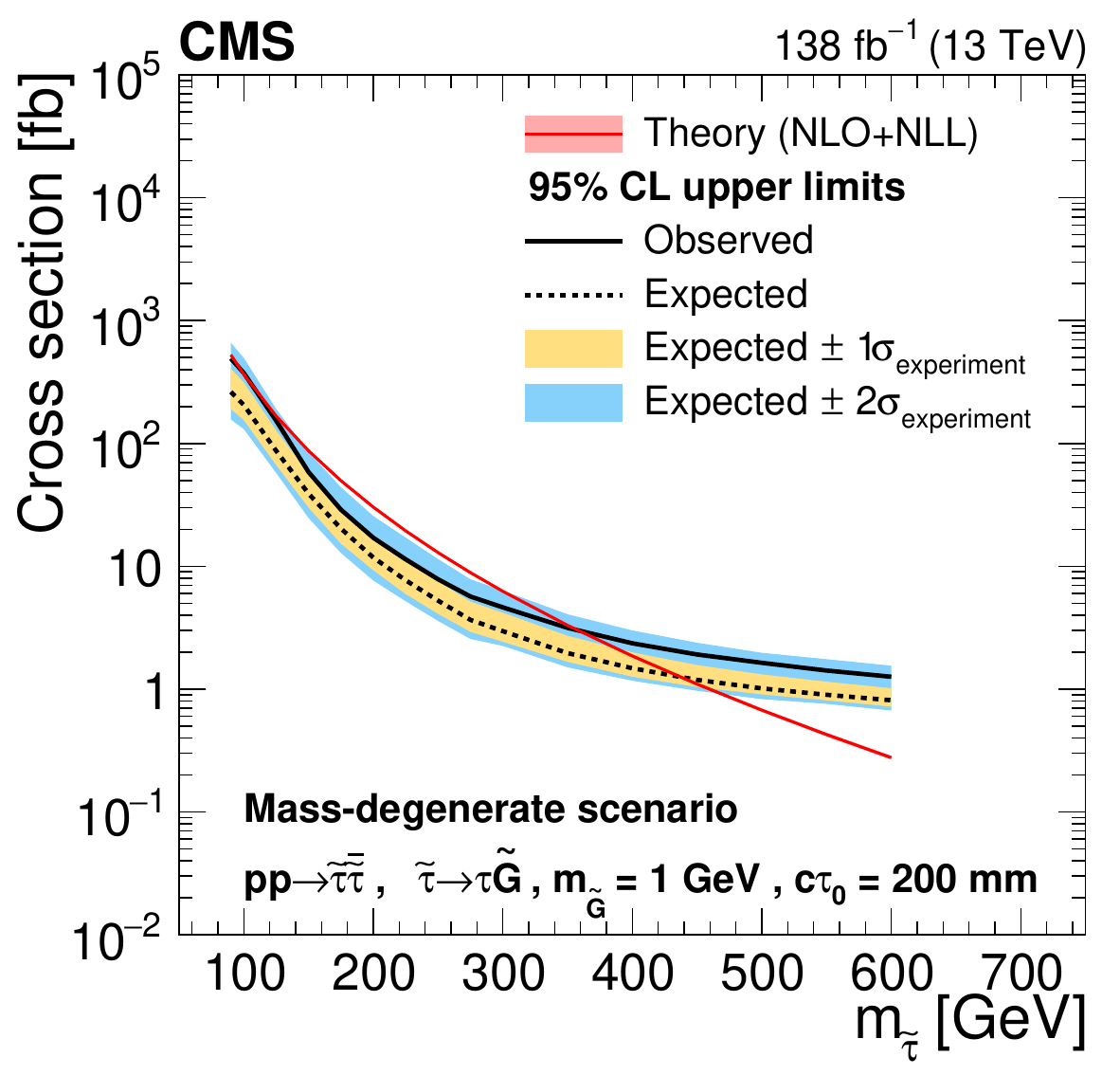}
    \includegraphics[width=0.42\linewidth]{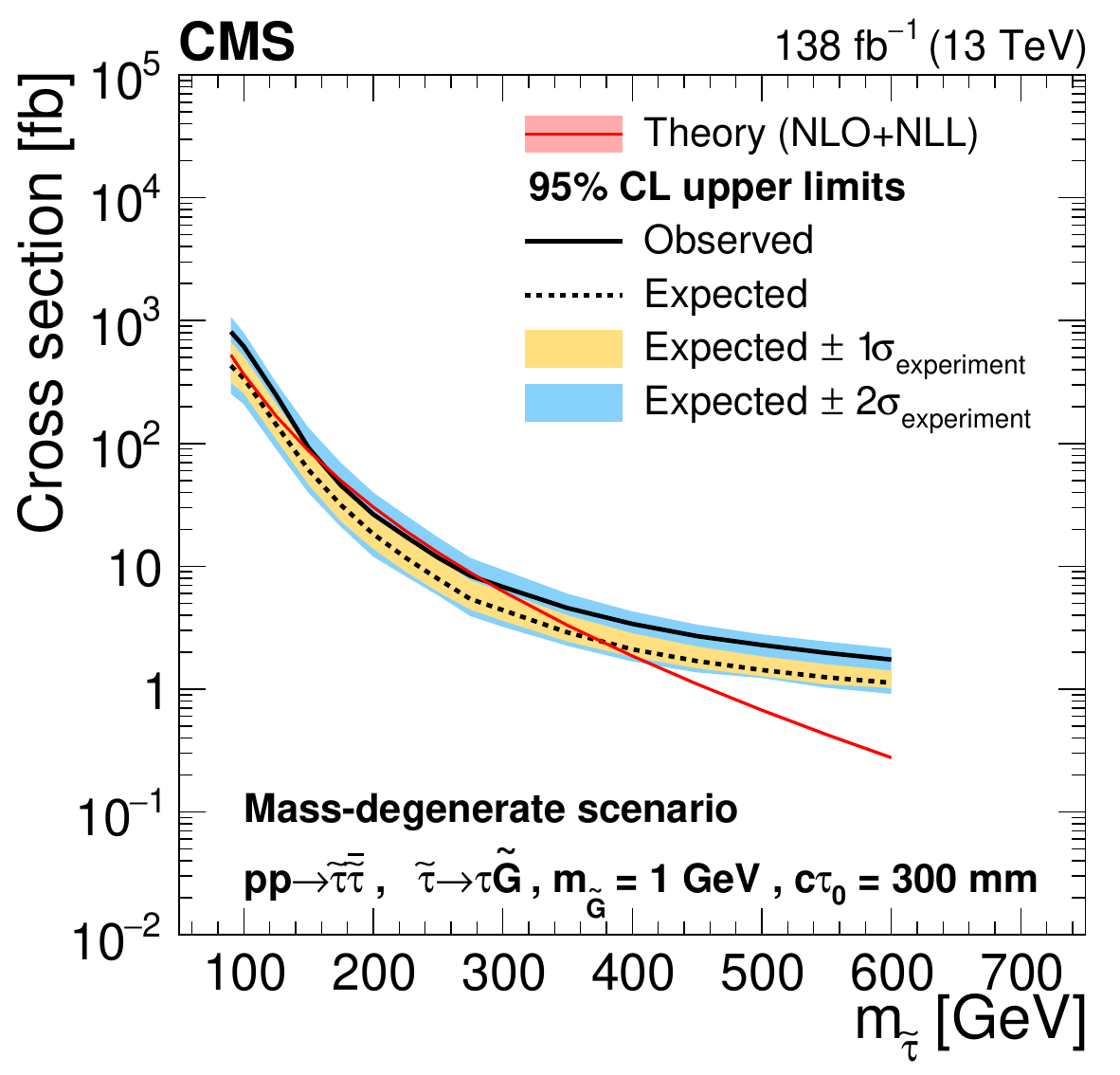}
    \caption{
    Cross section upper limits at 95\% \CL for the pair production of long-lived staus decaying to a \PGt lepton and a nearly massless gravitino ($m_\PXXSG=1$\GeV) in the mass-degenerate scenario, as a function of $\mstau$, for $\ctauz=$ 10, 30, 50, 100, 200, and 300\mm (left to right, upper to lower).
    The inner (yellow) and outer (blue) bands indicate the regions containing 68 and 95\%, respectively, of the distribution of limits expected under the background-only hypothesis.
    The central values of the expected upper limits are denoted by the dashed black line.
    The solid black line represents the observed upper limits.
    The signal cross sections and uncertainties evaluated using NLO$+$NLL calculations are shown as a red line and shaded band.
    }
    \label{fig:limits_1d_md}
\end{figure}

The final results are obtained by simultaneously fitting all the SR bins from the 2016, 2017, and 2018 data sets, using the CMS statistical analysis and combination tool \textsc{Combine}~\cite{CMS:2024onh}.
Figure~\ref{fig:limits_2d} shows the observed and expected exclusion limits for the pair production of long-lived staus decaying to a \PGt lepton and a nearly massless gravitino ($m_\PXXSG=1$\GeV) in the plane of the stau mass and the proper decay length.
In the maximally mixed scenario, the observed (expected) range of excluded \mstau values is 126--260 (90--335)\GeV for $\ctauz = 50\mm$, and the observed (expected) range of excluded \ctauz values is 21--94 (12--169)\mm for $\mstau = 200\GeV$.
In the mass-degenerate scenario, the observed (expected) range of excluded \mstau values is 90--425 (90--491)\GeV for $\ctauz = 50\mm$, and the observed (expected) range of excluded \ctauz values is 6--333 (5.5--442)\mm for $\mstau = 200\GeV$.
The exclusion limits in the mass-degenerate scenario are stronger than in the maximally mixed scenario, as the former has a larger cross section from the production of both $\PSGtDo$ and $\PSGtDt$, as discussed in Section~\ref{sec:simulation}.
Upper limits on the cross section as functions of $\mstau$ are presented in Figs.~\ref{fig:limits_1d_mm} and~\ref{fig:limits_1d_md}, for maximally mixed and mass-degenerate scenarios, respectively.

As reported in Table~\ref{tab:yields_sr} and Fig.~\ref{fig:yields_sr}, the predicted background is lower than the observed yield by approximately one standard deviation in bin 6.
Since bin 6 corresponds to an intermediate range of \ptjt (50--100\GeV), high \ptmiss (${>}250\GeV$), and any value of \mTii, it is sensitive to a broad range of \mstau and \ctauz signal hypotheses.
This results in the observed limits being weaker than the expected ones by approximately one standard deviation, across all tested hypotheses in the plane of $\mstau$ and $\ctauz$.

The upper limits on the production cross section become weaker for \ctauz below about 10\mm, as the \distauh candidates have smaller displacements and resemble background jets.
For \ctauz above a few hundred millimeters, the stau decays in the outer layers of the tracker, which reduces the reconstruction efficiency of \distauh candidates as the charged pions from the decay of the $\PGt$ lepton have fewer hits in the tracker.
In addition, a stau with high \ctauz can produce a signature in the tracker similar to that of a \distauh candidate with very small $\abs{\dxy}$ or of a background jet.
These effects result in a degradation of the upper limits at large values of \ctauz.
With growing \mstau, there is a gain in signal event yields because of an increase in the momenta of the \distauh candidates and the \ptmiss in the events, which causes the upper limits to become more stringent.

The improvement in observed upper limits on the production cross section for $\mstau=300\GeV$ in the mass-degenerate scenario ranges from a factor of 5.7 at $\ctauz = 3\mm$ to 2.5 at $\ctauz = 300\mm$, compared to the best available limits at $\sqrt{s}=13\TeV$ reported in the search for long-lived staus in Ref.~\cite{ATLAS:2020wjh}.
The improvement in the observed (expected) exclusion limits is also significant, for example, the maximum excluded \sTau mass reaches 425\GeV (490\GeV) at $\ctauz = 50\mm$, compared to 330\GeV (315\GeV) in Ref.~\cite{ATLAS:2020wjh}.
The range of excluded \sTau masses in this search is similar to that in Ref.~\cite{ATLAS:2024vnc}, which also includes data collected at $\sqrt{s}=13.6\TeV$.
However, the range of excluded \ctauz values at low \sTau masses in Ref.~\cite{ATLAS:2024vnc} is wider (3--1200\mm at $\mstau = 100\GeV$) than that in the present search (6--200\mm at $\mstau = 100\GeV$) owing to the usage of novel triggers for reconstructing displaced tracks with low momentum.
The results are also complementary to some previous searches, \eg the search described in Ref.~\cite{CMS:2022syk}, which explored small lifetimes ($\ctauz < 2.5\mm$) for the maximally mixed scenario, and the one described in Ref.~\cite{ATLAS:2025fdm}, which targeted long lifetimes ($\ctauz > 900\mm$) for the mass-degenerate scenario.

\clearpage

\section{Summary}
\label{sec:summary}

A search for the direct pair production of long-lived superpartners of the tau lepton (staus) has been performed in final states with two hadronically decaying tau leptons (\tauh) using data collected by the CMS detector in 2016--2018, corresponding to an integrated luminosity of 138\fbinv.
The potentially large background from misidentified jets in the fully hadronic final state is efficiently suppressed by the \distau neural network algorithm, specifically designed to identify displaced \tauh candidates.
In the maximally mixed (mass-degenerate) scenario, stau masses, $\mstau$, in the 126--260 (90--425)\GeV range are excluded for a proper decay length, $\ctauz$, of 50\mm, and stau proper decay lengths in the range 21--94 (6--333)\mm are excluded for $\mstau = 200\GeV$.
The development of a dedicated algorithm targeting displaced \tauh signatures has led to a significant improvement in the experimental sensitivity.
These results improve the exclusion limits compared to previous searches~\cite{ATLAS:2019gti,CMS:2019eln,ATLAS:2020wjh,CMS:2022syk,CMS:2021kdm,ATLAS:2024fub,ATLAS:2025fdm} at $\sqrt{s}=13\TeV$, and extend the parameter space explored in the context of supersymmetry.

\begin{acknowledgments}
We congratulate our colleagues in the CERN accelerator departments for the excellent performance of the LHC and thank the technical and administrative staffs at CERN and at other CMS institutes for their contributions to the success of the CMS effort. In addition, we gratefully acknowledge the computing centers and personnel of the Worldwide LHC Computing Grid and other centers for delivering so effectively the computing infrastructure essential to our analyses. Finally, we acknowledge the enduring support for the construction and operation of the LHC, the CMS detector, and the supporting computing infrastructure provided by the following funding agencies: SC (Armenia), BMBWF and FWF (Austria); FNRS and FWO (Belgium); CNPq, CAPES, FAPERJ, FAPERGS, and FAPESP (Brazil); MES and BNSF (Bulgaria); CERN; CAS, MoST, and NSFC (China); MINCIENCIAS (Colombia); MSES and CSF (Croatia); RIF (Cyprus); SENESCYT (Ecuador); ERC PRG, TARISTU24-TK10 and MoER TK202 (Estonia); Academy of Finland, MEC, and HIP (Finland); CEA and CNRS/IN2P3 (France); SRNSF (Georgia); BMFTR, DFG, and HGF (Germany); GSRI (Greece); NKFIH (Hungary); DAE and DST (India); IPM (Iran); SFI (Ireland); INFN (Italy); MSIT and NRF (Republic of Korea); MES (Latvia); LMTLT (Lithuania); MOE and UM (Malaysia); BUAP, CINVESTAV, CONACYT, LNS, SEP, and UASLP-FAI (Mexico); MOS (Montenegro); MBIE (New Zealand); PAEC (Pakistan); MES, NSC, and NAWA (Poland); FCT (Portugal);  MESTD (Serbia); MICIU/AEI and PCTI (Spain); MOSTR (Sri Lanka); Swiss Funding Agencies (Switzerland); MST (Taipei); MHESI (Thailand); TUBITAK and TENMAK (T\"{u}rkiye); NASU (Ukraine); STFC (United Kingdom); DOE and NSF (USA).

\hyphenation{Rachada-pisek} Individuals have received support from the Marie-Curie program and the European Research Council and Horizon 2020 Grant, contract Nos.\ 675440, 724704, 752730, 758316, 765710, 824093, 101115353, 101002207, 101001205, and COST Action CA16108 (European Union); the Leventis Foundation; the Alfred P.\ Sloan Foundation; the Alexander von Humboldt Foundation; the Science Committee, project no. 22rl-037 (Armenia); the Fonds pour la Formation \`a la Recherche dans l'Industrie et dans l'Agriculture (FRIA) and Fonds voor Wetenschappelijk Onderzoek contract No. 1228724N (Belgium); the Beijing Municipal Science \& Technology Commission, No. Z191100007219010, the Fundamental Research Funds for the Central Universities, the Ministry of Science and Technology of China under Grant No. 2023YFA1605804, the Natural Science Foundation of China under Grant No. 12535004, and USTC Research Funds of the Double First-Class Initiative No.\ YD2030002017 (China); the Ministry of Education, Youth and Sports (MEYS) of the Czech Republic; the Shota Rustaveli National Science Foundation, grant FR-22-985 (Georgia); the Deutsche Forschungsgemeinschaft (DFG), among others, under Germany's Excellence Strategy -- EXC 2121 ``Quantum Universe" -- 390833306, and under project number 400140256 - GRK2497; the Hellenic Foundation for Research and Innovation (HFRI), Project Number 2288 (Greece); the Hungarian Academy of Sciences, the New National Excellence Program - \'UNKP, the NKFIH research grants K 131991, K 133046, K 138136, K 143460, K 143477, K 146913, K 146914, K 147048, 2020-2.2.1-ED-2021-00181, TKP2021-NKTA-64, and 2025-1.1.5-NEMZ\_KI-2025-00004 (Hungary); the Council of Science and Industrial Research, India; ICSC -- National Research Center for High Performance Computing, Big Data and Quantum Computing, FAIR -- Future Artificial Intelligence Research, and CUP I53D23001070006 (Mission 4 Component 1), funded by the NextGenerationEU program (Italy); the Latvian Council of Science; the Ministry of Education and Science, project no. 2022/WK/14, and the National Science Center, contracts Opus 2021/41/B/ST2/01369, 2021/43/B/ST2/01552, 2023/49/B/ST2/03273, and the NAWA contract BPN/PPO/2021/1/00011 (Poland); the Funda\c{c}\~ao para a Ci\^encia e a Tecnologia, grant CEECIND/01334/2018 (Portugal); the National Priorities Research Program by Qatar National Research Fund;  MICIU/AEI/10.13039/501100011033, ERDF/EU, ``European Union NextGenerationEU/PRTR", and Programa Severo Ochoa del Principado de Asturias (Spain); the Chulalongkorn Academic into Its 2nd Century Project Advancement Project, the National Science, Research and Innovation Fund program IND\_FF\_68\_369\_2300\_097, and the Program Management Unit for Human Resources \& Institutional Development, Research and Innovation, grant B39G680009 (Thailand); the Kavli Foundation; the Nvidia Corporation; the SuperMicro Corporation; the Welch Foundation, contract C-1845; and the Weston Havens Foundation (USA).
\end{acknowledgments}
\bibliography{auto_generated} 
\cleardoublepage \appendix\section{The CMS Collaboration \label{app:collab}}\begin{sloppypar}\hyphenpenalty=5000\widowpenalty=500\clubpenalty=5000\cmsinstitute{Yerevan Physics Institute, Yerevan, Armenia}
{\tolerance=6000
A.~Hayrapetyan, V.~Makarenko\cmsorcid{0000-0002-8406-8605}, A.~Tumasyan\cmsAuthorMark{1}\cmsorcid{0009-0000-0684-6742}
\par}
\cmsinstitute{Institut f\"{u}r Hochenergiephysik, Vienna, Austria}
{\tolerance=6000
W.~Adam\cmsorcid{0000-0001-9099-4341}, L.~Benato\cmsorcid{0000-0001-5135-7489}, T.~Bergauer\cmsorcid{0000-0002-5786-0293}, M.~Dragicevic\cmsorcid{0000-0003-1967-6783}, P.S.~Hussain\cmsorcid{0000-0002-4825-5278}, M.~Jeitler\cmsAuthorMark{2}\cmsorcid{0000-0002-5141-9560}, N.~Krammer\cmsorcid{0000-0002-0548-0985}, A.~Li\cmsorcid{0000-0002-4547-116X}, D.~Liko\cmsorcid{0000-0002-3380-473X}, M.~Matthewman, J.~Schieck\cmsAuthorMark{2}\cmsorcid{0000-0002-1058-8093}, R.~Sch\"{o}fbeck\cmsAuthorMark{2}\cmsorcid{0000-0002-2332-8784}, M.~Shooshtari\cmsorcid{0009-0004-8882-4887}, M.~Sonawane\cmsorcid{0000-0003-0510-7010}, W.~Waltenberger\cmsorcid{0000-0002-6215-7228}, C.-E.~Wulz\cmsAuthorMark{2}\cmsorcid{0000-0001-9226-5812}
% ...
\par}
\vskip\cmsinstskip
\dag:~Deceased\\
$^{1}$Also at Yerevan State University, Yerevan, Armenia\\
$^{2}$Also at TU Wien, Vienna, Austria\\
% ...
\end{sloppypar}
%%% END EDITABLE REGION %%%
% skeleton_end
\end{document}